\documentclass[a4paper,fleqn]{cas-dc}

\usepackage{enumitem}
\usepackage{lipsum}
\usepackage[utf8]{inputenc}
\usepackage[numbers]{natbib}
\usepackage{graphicx}
\usepackage{pifont}
\usepackage{amssymb}
\usepackage{wasysym}
\usepackage{longtable}
\usepackage{booktabs} 
\usepackage{ltablex}  
\usepackage{todonotes}
\usepackage{comment}
\usepackage{enumitem}
\usepackage{float}
\usepackage{stfloats}     
\usepackage[section]{placeins}
\usepackage{subcaption} 
\usepackage{xcolor}
\colorlet{blue}{black}
\usepackage{tcolorbox}
\usepackage{hyperref}
\usepackage[table]{xcolor}
\usepackage{fontawesome}
\usepackage{rotating}
\usepackage{adjustbox}
\usepackage{placeins}
\usepackage{caption}
\DeclareUnicodeCharacter{2212}{\textminus}
\usepackage{cleveref}

\def\tsc#1{\csdef{#1}{\textsc{\lowercase{#1}}\xspace}}
\tsc{CE}
\tsc{OSS}

\ExplSyntaxOn
\cs_set:Npn \__first_footerline:
{
  \group_begin:
  \small
  \sffamily
  \ifnum\theblind>0\relax
  \else
    \__short_authors:
  \fi
  \group_end:
}
\ExplSyntaxOff

\begin{document}
\let\WriteBookmarks\relax
\def\floatpagepagefraction{1}
\def\textpagefraction{.001}

\shorttitle{Chaos Engineering in the Wild: Findings from GitHub}

\shortauthors{Owotogbe et al.}

\title[mode=title]{Chaos Engineering in the Wild: Findings from GitHub}                   

\tnotemark[1]
\tnotetext[1]{This work presents a large-scale empirical study of the adoption, evolution, and practical use of chaos engineering tools in open-source GitHub repositories.}



\author[1,2]{Joshua Owotogbe}
\cormark[1]
\ead{j.s.owotogbe@tilburguniversity.edu}

\author[1,2]{Indika Kumara}
\ead{i.p.k.weerasinghadewage@tilburguniversity.edu}

\author[4]{Dario Di Nucci}
\ead{ddinucci@unisa.it}

\author[1,3]{Damian Andrew Tamburri}
\ead{datamburri@unisannio.it}

\author[1,2]{Willem-Jan van den Heuvel}
\ead{w.j.a.m.vdnHeuvel@tilburguniversity.edu}


\affiliation[1]{
    organization={Jheronimus Academy of Data Science},
    city={'s-Hertogenbosch},
    state={North Brabant},
    country={Netherlands}
}

\affiliation[2]{
    organization={Tilburg University},
    city={Tilburg},
    state={North Brabant},
    country={Netherlands}
}

\affiliation[3]{
    organization={University of Sannio},
    city={Benevento},
    state={Benevento},
    country={Italy}
}

\affiliation[4]{
    organization={University of Salerno},
    addressline={Via Giovanni Paolo II, 132},
    city={Fisciano},
    state={Salerno},
    postcode={84084},
    country={Italy}
}

\newcommand{\etal}{\textit{et al.}}



\begin{abstract}
Chaos Engineering (CE) enhances system resilience by deliberately injecting faults to uncover weaknesses before they cause outages. Despite the availability of many CE tools, little is known about how they are adopted and maintained in open-source software (OSS) projects. This study empirically characterizes the adoption, evolution, and practical use of CE tools in OSS projects, examining who adopts them, how they are maintained, and which fault types they target. We conducted a large-scale mining study of GitHub repositories associated with ten widely used CE tools. Starting from 5,410 candidate repositories, we systematically filtered and manually validated 1,275 records and analyzed repository metadata, commit histories, and documentation. We found that adoption is concentrated around a few tools, with Toxiproxy, Chaos Mesh, and Chaos Monkey accounting for 68.86\% of the validated repositories. In terms of adopter context, development is the predominant repository purpose (56.55\%), while industry represents the largest ownership category (34.82\%), closely followed by personal repositories (33.49\%). At the project level, activity varies substantially, with 51.42\% of repositories having at most 50 commits. In terms of fault coverage, network faults (44.85\%) and instance termination (29.96\%) together account for 74.81\% of the 2,410 observed fault instances, whereas application-level faults account for only 2.57\%. Taken together, these findings suggest that practitioners should consider repository activity and fault coverage when selecting CE tools, while researchers should investigate the factors behind concentrated adoption and limited application-level experimentation.
\end{abstract}

\begin{keywords}
Chaos Engineering, Resilience Testing, Fault Injection, Mining Software Repositories, Tool Adoption
\end{keywords}

\maketitle

\section{Introduction}

Modern software systems increasingly support business processes and critical
applications through complex distributed architectures composed of
interconnected services deployed across heterogeneous
infrastructure~\citep{Patel22,Fawcett2020,Gremlin2023Tools}. As these systems
grow in scale and complexity, they remain exposed to a range of failures,
including infrastructure faults such as node termination and network
corruption, as well as application faults such as uncaught exceptions and
method-call delays
~\citep{al2024exploring,Bairyev2023,Nombela23,Roa2022,chen2022big}.
Consequently, ensuring dependability and business continuity requires not only mechanisms for detecting failures, but also confidence that systems can
continue operating or recover gracefully when failures occur
~\citep{fogli2023chaos,torkura2020cloudstrike,malik2023chess}. Chaos
Engineering (CE) addresses this need by deliberately injecting faults into
production or pre-production environments and observing how systems behave
under adverse conditions
~\citep{Basiri2017,Green2023,siwach2022evaluating,jones2019resilience}.
Tools such as \textit{Chaos Monkey}~\citep{chaosmonkey},
\textit{Toxiproxy}~\citep{toxiproxy}, and
\textit{Chaos Mesh}~\citep{chaosmesh} operationalize this practice by
supporting faults such as instance termination, network disruption, and
resource exhaustion. Both academic and grey literature report growing
interest in and use of CE tools
~\citep{Gremlin23,IBM2024,KahilArenhage2021,jernberg2020getting,owotogbe2025chaos}.

\textcolor{blue}{
CE originated in industrial resilience practice, with Netflix publicly
documenting the use of \textit{Chaos Monkey} as part of its Simian Army in
2011~\citep{izrailevsky2011simian}. The practice was subsequently formalized
and examined across industrial and organizational settings
~\citep{basiri_chaos_2016,jernberg2020getting,gremlin2021state,
owotogbe2025chaos}. This evolution is particularly relevant as the broader
software engineering landscape continues to change. Recent advances in AI-assisted software engineering increasingly support development-time activities, including source-code analysis and generation,
testing, debugging, program repair, and vulnerability detection
~\citep{hou2024large,wang2024software}. In this context, CE provides a
complementary system- and runtime-level perspective by deliberately
introducing failure conditions and examining system behavior and recovery
~\citep{Basiri2017,fogli2023chaos,torkura2020cloudstrike}. Establishing how
CE is currently practiced therefore provides an empirical baseline for
understanding existing resilience engineering practices and how they may
evolve alongside advances in AI-assisted software engineering.
}

\textcolor{blue}{
Although CE adoption has already been examined empirically, existing evidence
is largely organization-centered, drawing on industrial experience reports,
organization-specific case studies, practitioner studies, and evaluations in
particular technical environments
~\citep{izrailevsky2011simian,basiri_chaos_2016,jernberg2020getting,
gremlin2021state,owotogbe2025chaos}. Such work provides important knowledge
about the introduction and operational use of CE within organizations.
However, comparatively less is known about CE from an open-source,
cross-project perspective. In particular, an integrated empirical account is still lacking of which CE tools are represented across public OSS repositories, how their
repository-visible adoption and development evolve over time, what kinds of
projects use them, how active those projects are, and which fault-injection
capabilities are configured or invoked in repository artifacts. This
constitutes a complementary knowledge gap rather than an absence of prior
empirical CE research. The distinction between organization-centered and repository-visible evidence is important because the two perspectives reveal different aspects of CE
practice. Organization-centered studies can directly investigate motivations,
operational decisions, and reported outcomes, whereas public repository
artifacts enable large-scale and longitudinal comparison across independently
maintained projects but cannot directly reveal those internal factors. We
therefore treat these evidence sources as complementary perspectives on CE
practice. Public GitHub repositories provide a suitable empirical setting for examining
the repository-visible perspective because they expose source code,
configuration files, documentation, commit histories, contributor
information, and repository metadata that can provide evidence of CE-tool
integration and project evolution over time. Repository mining has similarly
been used to investigate the adoption, evolution, and use of
software-engineering tools and practices in prior empirical studies
~\citep{sanchez2022mutation,ustunboyacioglu2024data,tutko2022software,
pickerill2020phantom,di2025investigating,choi2025technology}. Such evidence
can complement existing industrial knowledge by revealing cross-project
patterns that are difficult to observe through individual organizational
studies.}

\textcolor{blue}{
Building on this methodological foundation, we conduct a large-scale
repository-mining study to characterize CE-tool adoption and evolution, the
contexts of adopting repositories, their development activity and community
engagement, and the fault-injection practices represented in their repository
artifacts. Accordingly, we formulate the following five research questions:
}

\begin{description}

    \item[RQ1.] \textit{How are CE tools adopted in open-source projects?}
    RQ1 examines the extent and temporal trends of CE-tool adoption across
    GitHub repositories, including the distribution of validated
    repository-tool records across the selected CE tools and their evolution
    over time.

    \item[RQ2.] \textit{How has the GitHub activity of the top-10 CE tools
    evolved?}
    RQ2 examines the development trajectories of the selected CE tools by
    analyzing the histories of their official \emph{CE tool repositories},
    including development activity, release histories, and community
    participation.

    \item[RQ3.] \textit{What open-source projects adopt CE tools?}
    RQ3 characterizes the contexts in which CE tools are used by examining
    repository purpose, including software development, research, teaching,
    and learning, together with repository ownership/origin, including
    industry, academia, public organizations, and personal repositories.

    \item[RQ4.] \textit{How active are the open-source projects that adopt CE
    tools, and what evidence reflects their prominence in the developer
    community?}
    RQ4 examines the development activity and community engagement of
    repositories associated with the selected CE tools using indicators such
    as commits, contributors, stars, forks, and watchers.

    \item[RQ5.] \textit{Which capabilities of CE tools are exploited by
    open-source projects?}
    RQ5 examines the fault-injection practices represented in the validated
    repositories by identifying the fault types configured or invoked in
    repository artifacts and mapping them to a common fault taxonomy.

\end{description}

\textcolor{blue}{
Together, these research questions follow two complementary analytical perspectives: a \textbf{tool-level} perspective (RQ1 and RQ2) that examines CE tool evolution and maturity, and a \textbf{repository-level} perspective (RQ3 to RQ5) that analyzes adoption and use across open-source projects. The perspectives are methodologically linked, as insights from tool evolution inform the analysis of adoption and practice, ensuring coherence across data collection, analysis, and research questions.}

We first identified ten widely used CE tools and analyzed 1,275 repositories selected from 5,410 candidates after removing duplicates and false positives. Adoption rose substantially after 2018, whereas the introduction of new CE tools peaked in 2018 and subsequently shifted toward refinement and adoption. Toxiproxy~\citep{toxiproxy}, Chaos Mesh~\citep{chaosmesh}, and Chaos Monkey~\citep{chaosmonkey} dominated usage, accounting for 68.86\% of repositories. Most repositories supported development (56.5\%), followed by teaching (10.4\%), learning (12.7\%), and research (6.7\%), with industry ownership prevailing (34.8\%). Activity varied widely: 51.4\% had at most 50 commits, 14.5\% had more than 1,000 commits, and highly active repositories were predominantly development-oriented. CE experiments mainly targeted network faults (44.9\%), followed by instance termination (30.0\%), resource stress (22.6\%), and application faults (2.6\%).

This study bridges CE research and practice by providing empirical evidence on how CE tools are developed, adopted, and used in real systems. The results guide practitioners by identifying widely used and well-maintained tools, supporting informed selection and integration into resilience workflows, and documenting CE adoption in industry and open-source projects. It also characterizes current CE practice and outlines future research directions, including tool evolution, adoption factors, and extensions to domains such as data and machine-learning pipelines.

The remainder of this paper is structured as follows. \Cref{sec:relatedwork} summarizes previous research on CE tools and highlights the gaps that motivate this study, and outlines the study’s research contribution and novelty. \Cref{sec:Method} outlines the study methodology, which details the data collection and analysis procedures. \Cref{sec:result} presents our findings, while \Cref{sec:discussion} explores the implications for researchers and practitioners. \Cref{sec:threats} discusses threats to the validity of our study.
Finally, \Cref{sec:conclusion} concludes the paper and offers recommendations for future work.

\section{Related Work and Research Contributions}
\label{sec:relatedwork}

This section reviews relevant academic, industrial, and practitioner evidence
on Chaos Engineering (CE), highlighting its objectives, empirical scope, and
limitations to identify the gap addressed in this work. It then presents the
research contributions and novelty of this study.

\subsection{Related Work}

Prior CE research can be organized into three complementary strands:
\emph{foundational and adoption-oriented studies}, \emph{tools and empirical
evaluations}, and \emph{studies synthesizing CE practice}.

\textbf{CE foundations and organization-centered adoption evidence.}

Early documented CE practice emerged from industrial resilience work at Netflix. Netflix publicly documented its Simian Army in 2011, including \textit{Chaos Monkey}, which deliberately disabled production instances to assess whether services could tolerate
infrastructure failures~\citep{izrailevsky2011simian}. The practice was later
formalized by \citet{basiri_chaos_2016}, who articulated CE as controlled
experimentation for assessing the resilience of distributed systems.

Subsequent work examined CE adoption more explicitly in organizational
settings. \citet{jernberg2020getting} developed an implementation framework
covering discovery, implementation, sophistication, and expansion, and
evaluated its initial stages in an industrial setting. This work provides
empirical guidance on introducing CE within an organization, but its scope is
organization-specific rather than cross-project. Broader practitioner
evidence is provided by Gremlin's \textit{State of Chaos Engineering 2021}
report~\citep{gremlin2021state}, which documents CE adoption, production use,
and commonly exercised experiment types across a wider practitioner
population. Together, these studies establish that CE adoption and use have
already been observed in industrial and organizational settings.

\textbf{CE tools, frameworks, and empirical evaluations.}

A second stream focuses on extending CE techniques and evaluating their
effects in specific technical environments.
\citet{torkura2020cloudstrike} extended CE to cloud security through
CloudStrike, while \citet{nikolaidis2021frisbee},
\citet{camacho2022chaos}, and \citet{chen2024microfi} developed frameworks
for Kubernetes-based benchmarking, hybrid-cloud experimentation, and
request-level fault injection, respectively. Complementing these
tool-oriented studies, \citet{al2024exploring} experimentally examined the
effects of injected faults on serverless and containerized applications.
Collectively, these studies demonstrate the applicability of CE in particular
systems and environments, but they do not characterize CE-tool adoption and
use across a broad population of independently maintained OSS repositories.

\textbf{CE reviews and broader synthesis.}

At a broader level, \citet{owotogbe2025chaos} synthesized academic and grey
literature to characterize CE concepts, tools, practices, adoption
considerations, and fault-injection strategies. More recently,
\citet{owotogbe2026ten} examined the evolution of CE across academia and
industry, tracing research milestones, tooling developments, application
domains, and emerging directions. Together, these studies provide a broader
synthesis of how CE has developed and been reported in research and practice.
However, their primary objective is literature- and evolution-oriented
synthesis rather than a large-scale, cross-project analysis of adopter
repositories and their repository-visible practices.

\textbf{Organization-centered versus open-collaborative evidence.}

Prior CE research provides two complementary evidence settings.
Organization-centered studies capture adoption motivations, operational
contexts, and reported outcomes
~\citep{izrailevsky2011simian,basiri_chaos_2016,jernberg2020getting,
gremlin2021state}. In contrast, open-collaborative repository evidence
captures observable OSS artifacts and enables cross-project and longitudinal
analysis~\citep{sanchez2022mutation,ustunboyacioglu2024data,tutko2022software,
pickerill2020phantom,di2025investigating,choi2025technology}, but cannot
directly reveal organizational motivations or operational outcomes.
Our study complements prior organization-centered evidence by examining how
CE is represented across public OSS repositories.

\begin{table*}[!ht]
 \tiny
 \caption{Comparison of representative Chaos Engineering (CE) studies and their coverage of key analytical dimensions. 
 The comparison is based on each study’s \textbf{objective}, \textbf{methodology}, and whether it addresses major aspects of CE research, 
 including \textbf{adoption patterns (Adopt.)}, \textbf{tool evolution (Evol.)}, \textbf{adopter context (Ctx.)}, 
 \textbf{project activity and popularity (Act.)}, and \textbf{fault-injection capabilities (Faults)}.}
 \label{tab:ce_empirical_comparison}
 \begin{tabular}{llp{1.7cm}p{1.7cm}cccccp{2cm}}
 \hline
 \textbf{Study} & \textbf{Year} & \textbf{Objective / Focus} & \textbf{Methodology} & \textbf{Adopt.} & \textbf{Evol.} & \textbf{Ctx.} & \textbf{Act.} & \textbf{Faults} & \textbf{Findings / Limitation} \\
 \hline
 \cite{izrailevsky2011simian} & 2011 &
 Industrial production resilience testing using Netflix's Simian Army. &
 Industrial experience report (grey literature); production fault injection. &
 \LEFTcircle & \ding{55} & \ding{51} & \ding{55} & \ding{51} &
 Documents production use of fault injection at Netflix; organization-specific
 and does not examine cross-project OSS adoption. \\
 \hline

 \cite{basiri_chaos_2016} & 2016 &
 Formalization of Chaos Engineering principles and industrial resilience
 experimentation. &
 Peer-reviewed experience-based software-engineering article. &
 \LEFTcircle & \ding{55} & \ding{51} & \ding{55} & \LEFTcircle &
 Formalizes CE based on industrial experience; does not provide large-scale
 cross-project adoption evidence. \\

 \hline
 \cite{jernberg2020getting} & 2020 & Industrial framework for CE adoption and process maturity. & Design science; literature and tool survey; interviews; industrial validation. & \LEFTcircle & \LEFTcircle & \ding{51} & \ding{55} & \ding{51} & Structured CE adoption framework; validated in a single enterprise; no repository-level analysis. \\
 \hline
 \cite{torkura2020cloudstrike} & 2020 & Security Chaos Engineering via \textit{CloudStrike} and RDFI framework. & Tool development and experimental evaluation (AWS, GCP). & \ding{55} & \LEFTcircle & \ding{51} & \ding{55} & \ding{51} & Demonstrated CE for security; effective fault automation; limited to cloud infrastructure; lacks adoption analysis. \\
 \hline

 \cite{gremlin2021state} & 2021 & Practitioner evidence on CE adoption, production use, and experiment types. & Industry survey and platform evidence (grey literature). &
 \ding{51} & \ding{55} & \ding{51} & \ding{55} & \ding{51} &
 Provides broader practitioner evidence of CE use; does not examine
 cross-project OSS repository adoption. \\
 \hline
 \cite{nikolaidis2021frisbee} & 2021 & Introduced Frisbee, a Kubernetes-native benchmarking suite for CE-based recovery testing. & Tool implementation and experiments (TiKV, Redis). & \ding{55} & \ding{55} & \ding{51} & \ding{55} & \ding{51} & Automates CE-based benchmarking; academic prototype; no adoption or evolution study. \\
 \hline
 \cite{camacho2022chaos} & 2022 & Designed Pystol, a SPL-based CE platform; comparative analysis of 13 tools. & Tool design, SPL modeling, hybrid-cloud experiments (AWS/OpenShift). & \LEFTcircle & \ding{51} & \ding{51} & \LEFTcircle & \ding{51} & Formal CE architecture; open-source release; small-scale evaluation; no adoption or evolution analysis. \\
 \hline
 \cite{chen2024microfi} & 2024 & Proposed MicroFI, a fine-grained CE framework for request-level fault injection. & Experimental evaluation on open-source benchmarks (Hipster Shop, TrainTicket, Hotel Reservation). & \ding{55} & \LEFTcircle & \ding{51} & \LEFTcircle & \ding{51} & Improves fault-injection precision and automation; limited to controlled workloads and request-level faults. \\
 \hline
 \cite{al2024exploring} & 2024 & Empirical analysis of CE’s impact on dependability under variable user loads. & Exploratory experiments on AWS Lambda and containerized microservices (760 trials). & \ding{55} & \ding{55} & \ding{51} & \ding{55} & \ding{51} & Showed CE exposes performance bottlenecks; delay faults only; no adoption or tool analysis. \\
 \hline
 \cite{owotogbe2025chaos} & 2025 & \textit{Chaos Engineering: A Multi-Vocal Literature Review.} Unified theory and taxonomy of CE. & Multivocal Literature Review (96 sources: 49 academic, 47 grey). & \LEFTcircle & \ding{51} & \ding{51} & \LEFTcircle & \ding{51} & Unified CE concepts and adoption taxonomy; qualitative only, no quantitative validation. \\
 \hline
 \cite{owotogbe2026ten} & 2026 &
 Longitudinal synthesis of CE evolution across research, industry practice,
 tooling, and emerging directions. &
 Curated synthesis of 103 peer-reviewed studies, 12 milestone studies, and
 open-source tooling evidence. &
 \LEFTcircle & \ding{51} & \LEFTcircle & \LEFTcircle & \ding{51} &
 Traces CE evolution across academia and industry; integrates prior GitHub
 tooling evidence but does not provide a cross-project analysis of adopter
 repositories. \\
 \hline
 \textbf{Ours} & 2026 & Large-scale cross-project analysis of CE-tool adoption, evolution, adopter context, repository activity, and fault practices. & GitHub mining and quantitative analysis of 1,275 validated repositories across 10 CE tools. & \ding{51} & \ding{51} & \ding{51} & \ding{51} & \ding{51} & First large-scale, data-driven study of CE adoption; quantifies adoption, evolution, and fault practices. \\
 \hline
 \end{tabular}
 
 \begin{minipage}{\textwidth}
 \scriptsize{\textit{Note.} \ding{51} = explicitly addressed; \LEFTcircle = partially addressed; \ding{55} = not addressed.}
 \end{minipage}
\end{table*}

\textcolor{blue}{
As summarized in \Cref{tab:ce_empirical_comparison}, CE adoption and use have
already been examined through industrial reports, organization-specific
studies, practitioner evidence, and experimental work
~\citep{izrailevsky2011simian,basiri_chaos_2016,jernberg2020getting,
gremlin2021state,torkura2020cloudstrike,nikolaidis2021frisbee,
camacho2022chaos,chen2024microfi,al2024exploring}. Broader studies have also
synthesized CE practices and traced the evolution of the field across
academia and industry
~\citep{owotogbe2025chaos,owotogbe2026ten}. However, CE remains comparatively under-characterized from a large-scale, open-collaborative repository perspective. Existing evidence does not provide a cross-project account that jointly examines CE-tool representation and evolution, adopter contexts, repository activity, and fault-injection
practices across public OSS projects. Our study addresses this complementary
gap through large-scale GitHub repository mining, extending existing
industrial and organizational knowledge with repository-visible evidence.
}

\subsection{Research Contribution and Novelty}
\label{sec:Knowledgebody}

Building on the identified research gap, this study presents the first large-scale empirical analysis of CE tools using open-source repositories. We analyze 1,275 validated CE-adopting repositories covering ten widely used CE tools to examine adoption, evolution, and usage patterns, addressing limitations of prior conceptual and case-based work. The six contributions link these findings to the analytical dimensions defined by the five research questions.

\begin{enumerate}[label=\textbf{C\arabic*.},leftmargin=1.1cm]
    \item \textbf{Characterization of adoption patterns and trends (RQ1):}  
    We quantified the extent of CE tool adoption across 1,275 CE-adopting repositories and analyzed temporal adoption trends to identify which tools show sustained, increasing, or limited use over time.
    \item \textbf{Analyzing CE tool evolution and maturity (RQ2):}  
    We examined release activity, community participation, and repository maintenance to assess how widely used CE tools evolve.
    \item \textbf{Identifying CE adopter contexts and project types (RQ3):}  
    Categorization of projects by purpose (software development, research, teaching, learning) and ownership (industry, academia, personal, public) to map where and how CE is practiced.
    \item \textbf{Examination of project activity, popularity, and evolution (RQ4):}  
    We analyzed commits, contributors, releases, stars, forks, and watchers to evaluate the activity level and visibility of CE-adopting projects.
    \item \textbf{Investigation of fault-injection practices (RQ5):}  
    We examined which fault types (e.g, network faults, instance termination, application faults) are used in practice and how they reflect actual CE usage.
    \item \textbf{Open dataset and replication package (supporting RQ1–RQ5):}  
    We release a curated dataset of 1,275 CE-adopting repositories spanning ten CE tools, along with all scripts and metadata, to support replication and future research.
\end{enumerate}

Together, these contributions establish the first comprehensive, data-driven baseline for understanding how CE tools are adopted, evolved, and applied in real-world software projects and teaching/learning environments.

\section{Research Design}
\label{sec:Method}
\begin{figure*}[ht]
 \centering
  \includegraphics[width=0.55\textwidth]{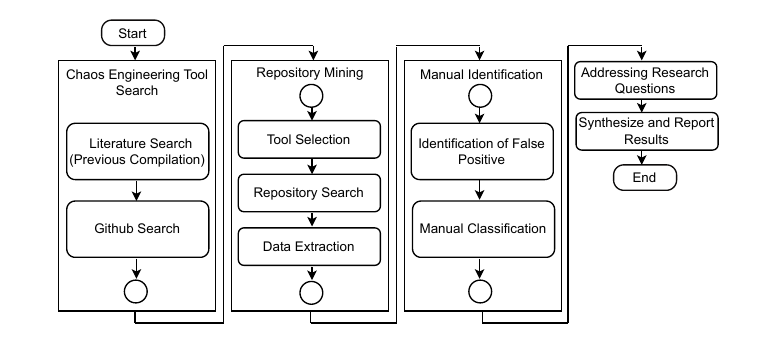}
 \caption{Data collection and analysis process based on \citet{sanchez2022mutation} and \citet{ustunboyacioglu2024data}.}
 \label{fig:CE_MLR_PROCESSS}
\end{figure*}

We employed a research methodology similar to prior studies with similar goals, e.g., analyzing the adoption of mutation testing~\citep{sanchez2022mutation} and data quality tools~\citep{ustunboyacioglu2024data} in GitHub projects. \Cref{fig:CE_MLR_PROCESSS} shows the key steps in our research design, including CE tool search, repository mining, manual identification, and the analysis steps used to address the research questions.  We completed all searches and data collection activities in August 2026.

\textcolor{blue}{
GitHub was selected because it provides the repository artifacts needed
to answer the research questions, including source and configuration files,
release histories, commits, contributor records, and repository metadata.
Specifically, these artifacts allow us to identify repositories that
integrate or invoke CE tools (RQ1), examine the development history of the
selected CE tools (RQ2), classify the purpose and ownership of adopting
repositories (RQ3), measure their activity and prominence (RQ4), and
identify fault types that are configured or invoked in repository artifacts
(RQ5). The study is therefore limited to evidence available in public GitHub
repositories and does not capture private repositories or projects hosted
on other platforms~\citep{kalliamvakou2016depth}.
}

\textcolor{blue}{
Given this scope, repository mining allows us to establish what is recorded
in repository artifacts, but not the motivations underlying those practices.
For example, the artifacts can show that a CE tool is integrated or invoked,
which fault types are configured or invoked, and how repositories change over
time. They cannot establish why developers or organizations selected a
particular CE tool or adopted a particular CE practice; such motivations
require complementary methods such as interviews or surveys ~\citep{stol2015theory}.
}

\textcolor{blue}{
Accordingly, we distinguish between two repository types throughout the
analysis. A \emph{CE tool repository} is the official GitHub repository in
which a selected CE tool is developed and maintained; these repositories
are analyzed in RQ2. In contrast, a \emph{CE-adopting repository} is an OSS
project repository for which manual inspection identified evidence that one
of the selected CE tools was integrated or invoked for a CE-related purpose;
these repositories are analyzed in RQ1 and RQ3-RQ5.
}

Data collection comprised three stages: identifying CE tools from prior literature and GitHub and analyzing their evolution, community engagement, and maintenance (RQ1 – RQ2); selecting the ten most frequently used tools from GitHub repositories (RQ3 – RQ4); and manually validating and classifying associated repositories to confirm relevance and usage contexts (RQ5). The following subsections describe these stages in detail.

\subsection{Chaos Engineering Tools Search}
\textcolor{blue}{
We identified CE tools through two complementary sources: the systematic
literature search conducted in our previous multivocal literature review and
a complementary GitHub repository search. The former provided tools documented
in academic and practitioner literature, while the latter captured additional
tools visible in open-source development.
}

\paragraph{Previous Tool Compilation}
\textcolor{blue}{
We began with the 40 CE tools identified in our previous multivocal
literature review (MLR)~\citep{owotogbe2025chaos}. Briefly, the MLR
systematically searched Scopus, IEEE Xplore, ACM Digital Library, and Web of
Science using piloted CE-specific search strings, supplemented by Google
Scholar and Google Search to capture additional academic and grey literature.
After duplicate removal, eligibility screening, and quality assessment, 96
sources (49 academic and 47 grey-literature sources) were included in the
final synthesis. CE tools reported in these sources were extracted and
classified, yielding the 40 tools that form the literature-derived baseline
for the present study.
}

\paragraph{Complementary GitHub Search}

During the study, we found that some widely used CE tools are popular on GitHub but absent from academic literature, a pattern also observed for mutation testing tools~\citep{sanchez2022mutation}. To capture these tools, we searched GitHub for \textit{chaos engineering} repositories sorted by stars, a common proxy for popularity~\citep{borges2016understanding}. 

The search returned 546 repositories; we reviewed the top 150 ranked by
stars, representing the most visible repositories returned by the search,
consistent with prior work that uses stars as an indicator of repository
popularity~\citep{borges2016understanding,sanchez2022mutation}.

For each of the 150 candidate repositories, we manually reviewed documentation, descriptions, and metadata to determine whether the repository qualified as a CE tool, defined as software that injects faults or orchestrates experiments to assess system resilience. This process identified 61 repositories as bona fide CE tools not covered by prior literature. Combined with the 40 tools from the literature-based corpus, this yielded a consolidated set of 101 distinct CE tools that formed the basis of our analysis.

Next, we refined the dataset using four exclusion criteria: tools for which no identifiable and publicly accessible GitHub repository could be located (14), repositories explicitly marked as archived on GitHub (5), insufficient documentation to verify CE functionality (34), and tools that did not satisfy our operational definition of chaos engineering (7). This removed 60 tools, leaving 41 eligible tools.

\textcolor{blue}{Recent commit activity was not used as an exclusion criterion. We distinguished limited recent development activity from explicit repository archival, consistent with prior repository-mining work~\citep{ustunboyacioglu2024data}. Thus, \textit{PowerfulSeal} remained eligible because its repository was publicly accessible, non-archived, and sufficiently documented to verify its CE functionality, although its last observed commit was in 2021. Following prior repository-mining studies~\citep{sanchez2022mutation,ustunboyacioglu2024data}, we selected the ten highest-ranked eligible tools based on GitHub stars to obtain a tractable set of prominent CE tools for detailed analysis. GitHub stars were used as an indicator of repository popularity and community interest rather than as a measure of recent maintenance activity.}

\paragraph{Repository Mining} 
To study the adoption and evolution of CE tools, we mined GitHub repositories with verifiable evidence of tool usage, such as configuration files, dependency declarations, or command invocations. We applied a two-phase mining process: first, we identified repositories referencing the ten selected CE tools using tailored search queries; second, we extracted repository metadata via the GitHub REST API~\cite{github_rest_api_docs} for subsequent analysis. The following subsections describe these phases in detail.

\subsection{Repository Search}
In this phase, we aimed to identify repositories that demonstrate use of the selected CE tools, particularly those integrated as libraries or components~\citep{ustunboyacioglu2024data,sanchez2022mutation}. To achieve this, we designed tool-specific search queries by thoroughly examining the documentation of each tool to understand how it integrates into projects. \Cref{tab:chaos_tools_results} shows the search queries. We follow GitHub's official guidelines for code search.\footnote{\url{https://docs.github.com/en/search-github/searching-on-github/searching-code}}

For example, the search query \texttt{chaoskube extension:yaml} targets YAML files that define Chaoskube configurations. Because Kubernetes uses YAML for configuration and Chaoskube randomly terminates pods, such files indicate Chaoskube usage in repositories. We used an automated script that authenticates with the GitHub API, performs code searches for each tool, processes paginated results, and extracts unique repository names and owners. Data collection was constrained by the GitHub API: access was limited to recently active or searchable repositories, files larger than 384 KB were excluded, and each query returned at most 1,000 results\footnote{\url{https://docs.github.com/en/rest/search?apiVersion=2022-11-28\#about-the-search-api}}. We addressed these constraints by partitioning searches with the size qualifier, which filters files by byte size and supports a tool-agnostic, general-purpose scraping approach~\citep{sanchez2022mutation,ustunboyacioglu2024data}.

\textcolor{blue}{
The tool-specific searches returned 5,410 candidate repository-tool records
associated with the 10 selected CE tools. The detailed search queries and
the number of records retrieved for each tool are summarized in
\Cref{tab:chaos_tools_results}. Because a search result did not necessarily
indicate genuine CE-tool use, all candidate records were subsequently
manually validated. Records containing only incidental references to a tool,
such as mentions in comments, dependency lists, links, or unrelated
configuration files, were excluded. The verification and filtering procedures
that produced the final validated dataset are described in the following
sections.
}

\begin{table}[ht]
\tiny
\caption{Search strings and results for CE tools. FP stands for false positives. w/o = without false positives.}
\label{tab:chaos_tools_results}
\centering

\begin{tabular}{llrr}
\hline

\textbf{Tool Name} &
\textbf{Search String} &
\textbf{\# Repos} &
\textbf{\# Repos w/o FP}
\\
\hline

Chaos Monkey &
\begin{tabular}[t]{@{}l@{}}
chaosmonkey extension:yml \\
chaosmonkey extension:json \\
chaosmonkey filename:py
\end{tabular}
& 748 & 199
\\
\hline

Toxiproxy &
\begin{tabular}[t]{@{}l@{}}
toxiproxy extension:yml \\
toxiproxy extension:py \\
toxiproxy extension:php \\
toxiproxy extension:json \\
toxiproxy filename:makefile
\end{tabular}
& 1,623 & 370
\\
\hline

Chaos Mesh &
\begin{tabular}[t]{@{}l@{}}
chaos-mesh extension:yaml \\
chaos-mesh extension:yml
\end{tabular}
& 696 & 309
\\
\hline

ChaosBlade &
\begin{tabular}[t]{@{}l@{}}
chaosblade extension:yaml \\
chaosblade filename:makefile \\
chaosblade filename:dockerfile
\end{tabular}
& 107 & 44
\\
\hline

LitmusChaos &
litmus extension:yml
& 793 & 132
\\
\hline

Kube-monkey &
kubemonkey extension:yaml
& 61 & 32
\\
\hline

Pumba &
\begin{tabular}[t]{@{}l@{}}
pumba extension:yml \\
pumba filename:dockerfile \\
pumba filename:makefile
\end{tabular}
& 194 & 42
\\
\hline

Chaos Toolkit &
\begin{tabular}[t]{@{}l@{}}
chaostoolkit extension:yaml \\
chaostoolkit extension:yml
\end{tabular}
& 1,041 & 81
\\
\hline

PowerfulSeal &
\begin{tabular}[t]{@{}l@{}}
powerfulseal extension:yml \\
powerfulseal filename:dockerfile
\end{tabular}
& 34 & 11
\\
\hline

Chaoskube &
chaoskube extension:yaml
& 113 & 55
\\
\hline

\textbf{Total}
&
&
\textbf{5,410}
&
\textbf{1,275}
\\
\hline

\end{tabular}

\end{table}

\paragraph{Data Extraction}
After compiling the list of repositories that may use CE tools, we retrieve detailed information for each repository identified in the previous step. Using the GitHub REST API~\cite{github_rest_api_docs}, a custom mining script automated the collection of repository-level information, including name, description, creation date, last update, and URL. The script also extracted owner metadata (e.g., username and affiliation), project attributes (e.g., primary language, license, topics, and issue tracking), and activity metrics (e.g., commits, contributors, stars, forks, and watchers, along with timestamps of the first and last commits). We exported all extracted data to Microsoft Excel spreadsheets for organization and analysis. This structured dataset provides an overview of the activity, visibility, and technical characteristics of open-source projects that incorporate CE tools.

\subsection{Manual Review and Classification}
After collecting and consolidating all metadata, we conducted a manual
review to ensure the reliability of the dataset and to contextualize the
use of CE tools in open-source projects. This process involved three key
activities: (1) identification of false positives, (2) classification of
verified repositories according to their purpose and ownership, and
\textcolor{blue}{(3) identification and classification of the fault types observed in the repositories.}

\paragraph{Identification of False Positives}

After compiling a list of repositories using the selected CE tools, we manually reviewed it to identify and remove false positives. First, we analyzed the tools' documentation to understand how they are applied in CE experiments. Next, we manually inspected repository artifacts, including configuration files, dependencies, documentation, and source code, to distinguish incidental references from evidence of CE-tool use. We retained a repository when these artifacts showed that a CE tool was integrated or invoked for a CE-related purpose, such as pod termination, disk failure, or latency simulation. We excluded repositories with only incidental tool-name references, as well as duplicates and forks. To reduce researcher bias, we applied the same predefined screening criteria across all ten tools. The first author performed the initial false-positive assessment, after which the second author reviewed the screening decisions against the predefined criteria. Cases with ambiguous or insufficient evidence were flagged for joint review, and additional remarks were recorded where necessary. The authors then re-examined these cases and resolved disagreements through discussion until consensus was reached, following the consensus-based manual review procedure used in comparable repository-mining studies ~\citep{sanchez2022mutation}. Because repository artifacts do not indicate whether a CE experiment ran at runtime, our findings characterize observable repository-level use rather than verified runtime execution. Following this process, 1,275 of the 5,410 candidate repositories were retained in the final dataset, as summarized in \Cref{tab:chaos_tools_results}.

\paragraph{Repository Classification}


To address RQ3, we manually classified repositories by origin
(public, academic, industry, and personal) and purpose (learning,
research, teaching, development, or unclassified), following comparable
repository-mining studies
~\citep{sanchez2022mutation,ustunboyacioglu2024data,de2022software}.
Classification was based on repository documentation, source code, and
owner information. The first author performed the initial classification,
and the second author reviewed all assignments against the predefined
criteria and repository evidence. Ambiguous cases were jointly re-examined
and resolved through discussion until consensus was reached
~\citep{sanchez2022mutation}. \Cref{sec:Repo_Class} provides the category definitions and examples. A total of 174 repositories lacked sufficient evidence for reliable purpose
classification and were therefore labeled \textit{Unclassified} rather than
discarded. They were excluded from analyses comparing repository purpose
categories in RQ3-RQ4, while retained in the dataset for completeness and
reproducibility. Preliminary checks showed that their exclusion did not alter
tool-usage distributions or relative rankings
~\citep{sanchez2022mutation,ustunboyacioglu2024data}.


\paragraph{Fault-Injection Classification}

To address RQ5, we reviewed the official documentation of each selected CE
tool to identify supported fault types, parameters, scope, and terminology,
which we normalized into the common taxonomy in \Cref{tab:fault_types}:
network, instance termination, resource stress, and application-level faults
with their subcategories. For each validated repository, we recorded a fault
only when repository artifacts provided evidence that it was configured or
invoked; faults were not inferred solely from tool capabilities. Because
repositories could exhibit multiple faults, coding was multi-label, and no
fault was assigned when evidence was insufficient. The first author performed
the initial classification, and the second author reviewed all assignments
against the repository evidence, tool documentation, and taxonomy. Ambiguous
cases were jointly re-examined and resolved through discussion until consensus
was reached, using consistent coding criteria across all ten tools and following
consensus-based practices in comparable repository-mining studies
~\citep{sanchez2022mutation}.

\textcolor{blue}{
\section{Results and Analysis}
}
\label{sec:result}

\textcolor{blue}{
This section presents the quantitative results and analyzes the observed
patterns for each research question.
}



\subsection{RQ1: How are CE Tools Adopted in Open-source Projects?}

\textcolor{blue}{
To address RQ1, we analyze 1,275 CE-adopting repositories with manually verified evidence of CE-tool integration or invocation. We excluded repositories with only incidental references to CE tools as false positives. \Cref{tab:chaos_tools_results} summarizes the tools, search strings, and repository counts before and after validation.} Following the methodology of prior repository-based studies~\cite{sanchez2022mutation, ustunboyacioglu2024data}, we focused on the most prominent tools to ensure methodological consistency and to base our analysis on representative, empirically verifiable data. Among these, \textit{Toxiproxy} stands out as the most widely adopted, appearing in 370 repositories, followed by \textit{Chaos Mesh} (309), \textit{Chaos Monkey} (199), \textit{LitmusChaos} (132), and \textit{Chaos Toolkit} (81). Less-adopted tools include \textit{Chaoskube} (55), \textit{ChaosBlade} (44), \textit{Kube-Monkey} (32), \textit{Pumba} (42), and \textit{PowerfulSeal} (11). This concentration of activity around a small number of tools reflects a trend often observed in empirical studies of software testing technologies, where a few mature tools dominate practical adoption.

\begin{figure*}[ht]
 \centering
 \includegraphics[width=0.55\textwidth]{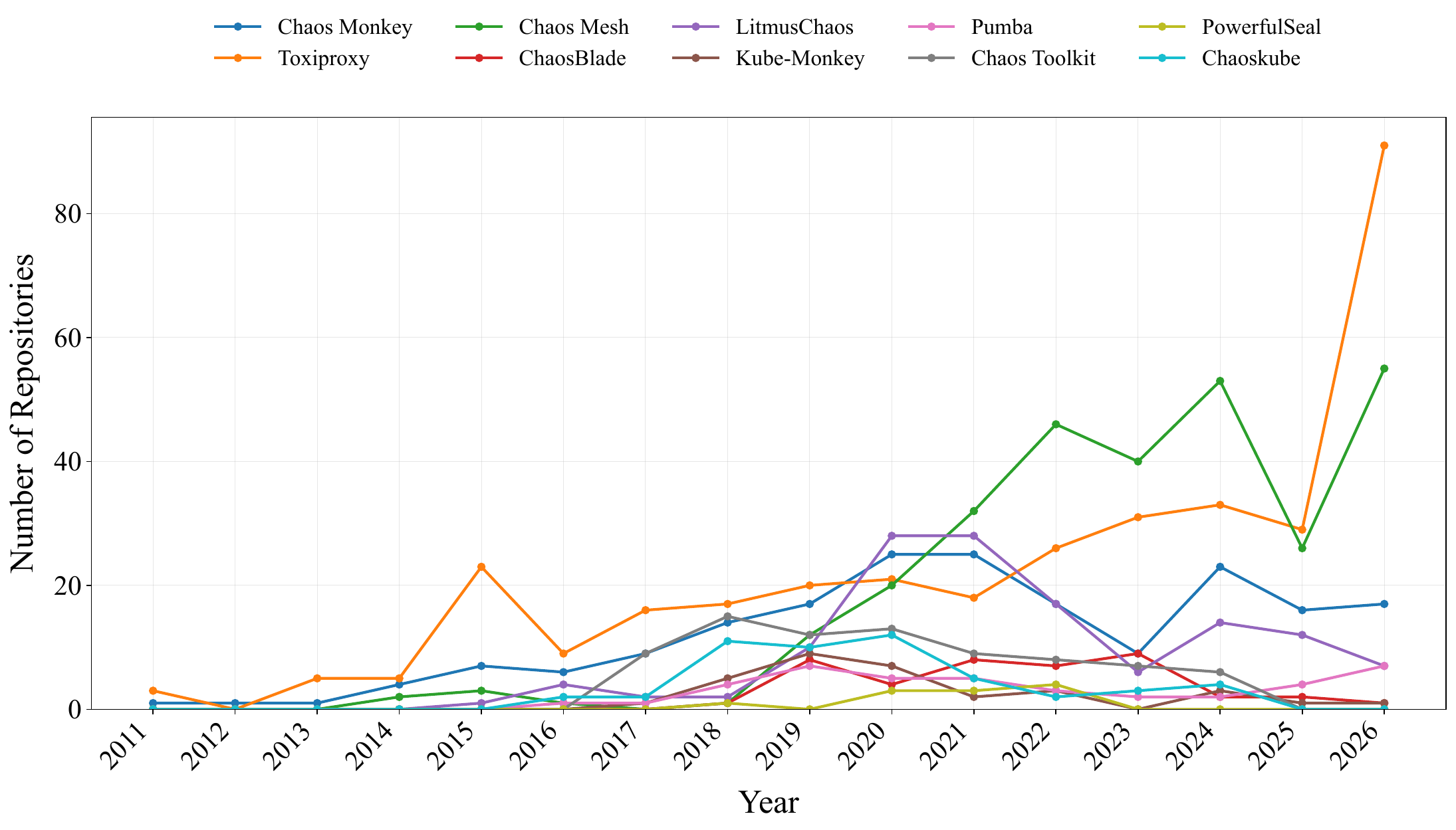}
 \caption{Yearly repository creation trends for the top-10 CE tools}
 \label{fig:CE_Yearly_Trends_Repo_new}
\end{figure*}

\Cref{fig:CE_Yearly_Trends_Repo_new} shows the yearly creation of repositories using each CE tool. Adoption increases steadily from around 2018, although trajectories vary across tools and years. Because data collection occurred in August 2026, the 2026 values represent only a partial year and should therefore be interpreted cautiously when comparing yearly trends. Adoption trajectories differ across tools. \textit{Toxiproxy} shows substantial growth over the study period, with a pronounced increase in 2026. \textit{Chaos Monkey}, an early CE framework, shows several growth phases, including a notable increase in 2024. Introduced in 2018, \textit{Chaos Mesh} saw rapid adoption after 2020, likely due to Kubernetes integration and support from the Cloud Native Computing Foundation (CNCF). \textit{LitmusChaos} follows a similar pattern, peaking around 2020 and declining thereafter. In contrast, \textit{ChaosBlade}, \textit{Kube-Monkey}, \textit{Pumba}, and \textit{Chaoskube} show comparatively lower and more variable adoption, while \textit{PowerfulSeal} remains minimally adopted. Commit histories indicate that many repositories adopt these tools long after their creation, a pattern also noted in prior studies~\citep{sanchez2022mutation,ustunboyacioglu2024data}. Overall, adoption concentrates on a small set of tools, particularly those with strong community support and cloud-native compatibility.

These trends align with broader shifts in the software ecosystem. The growth after 2018 coincides with widespread adoption of Kubernetes and the consolidation of DevOps and microservice architectures, which enabled CE integration~\citep{taibi2018continuous,offerman2022study,mitchell2023cloud}. By this period, DevOps automation was standard, and microservices provided isolation for resilience experiments. This pattern suggests that CE’s rise reflects the maturation of cloud-native engineering rather than transient interest in specific tools. Our repository analysis further shows that CE frameworks are integrated into projects through multiple mechanisms rather than a single standardized approach. Specifically, we identified three recurring integration modes. The first is \textit{configuration-driven integration}~\cite{xu2019configuration}, where experiments are defined in YAML manifests or Helm charts (common in \textit{Chaos Mesh} and \textit{LitmusChaos}). The second is \textit{container- or orchestration-level integration}~\cite{simonsson2021observability}, in which CE commands are invoked directly in Dockerfiles, CI/CD pipelines, or orchestration scripts (e.g., tools such as \textit{Pumba} and \textit{Chaos Monkey}). Finally, a smaller number of projects use \textit{API- or SDK-based integration} \cite{ofoeda2024api}, relying on programmatic interfaces provided by tools such as \textit{Chaos Toolkit} or \textit{ChaosBlade}. 

Integration styles mirror each tool’s platform focus: Kubernetes-native tools use configuration-driven approaches, platform-agnostic tools use container-level integration, and programmable environments use APIs. CE adoption in open-source projects remains uneven, with most activity concentrated in a few mature tools, particularly \textit{Toxiproxy}, \textit{Chaos Mesh}, \textit{Chaos Monkey}, and \textit{LitmusChaos}. The dominance of configuration- and automation-based integration reflects the tight coupling of CE with DevOps workflows and the growing embedding of resilience testing in cloud-native development.

\begin{tcolorbox}[colback=blue!3, colframe=black!50, boxrule=0.8pt, arc=1pt, left=4pt, right=4pt, top=2pt, bottom=2pt]

\textbf{RQ1:} 
Adoption of CE tools has increased since 2016 and accelerated after 2018, concentrating on a small set of mature tools-\textit{Toxiproxy}, \textit{Chaos Mesh}, \textit{Chaos Monkey}, and \textit{LitmusChaos}. These tools primarily integrate via configuration files and CI/CD pipelines, and their growth aligns with Kubernetes adoption and DevOps automation, reflecting the maturation of cloud-native software engineering.
\end{tcolorbox}

\subsection{RQ2: How has the GitHub Activity of the Top-10 CE Tools Evolved?}

\textcolor{blue}{
Understanding CE-tool evolution requires considering both how the tool landscape expands and how established tools continue to develop. We therefore examine these two complementary perspectives through the introduction of
\emph{new CE tools} and the version releases and repository activity of \emph{existing CE tools}.} Throughout this section, a \emph{new tool} refers to the initial introduction of a distinct CE tool, whereas a \emph{version
release} refers to a tagged release of an existing tool. Following \Cref{sec:Method}, we analyzed the ten selected CE tools using their version-release histories, commit timelines, and GitHub metrics, including watchers, stars, forks, commits, issues, and tagged version releases. We analyzed these data to characterize their development, maintenance, and community activity over time.

\begin{figure*}[ht]
 \centering
 \includegraphics[width=0.45\textwidth]{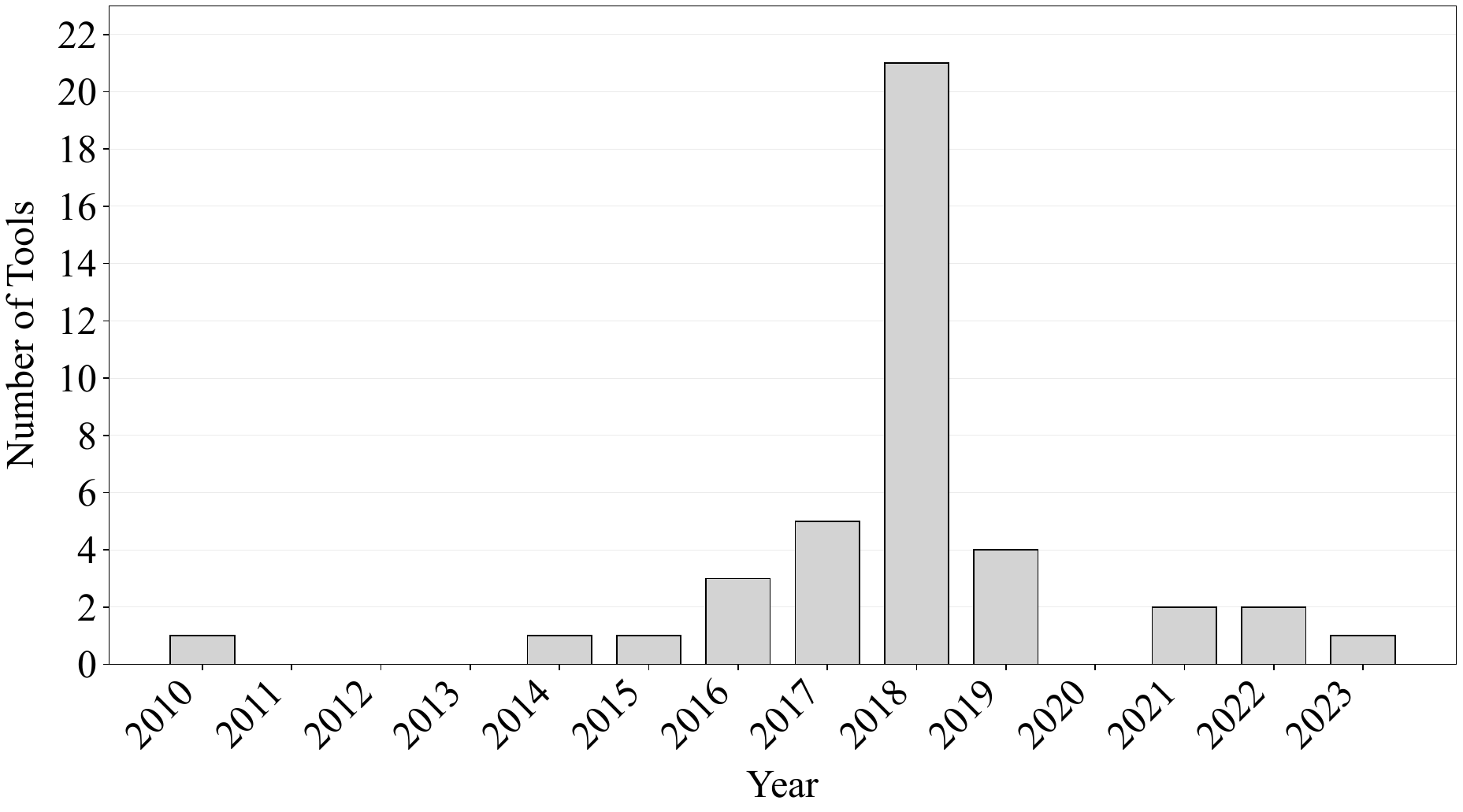}
 \caption{Number of distinct CE tools introduced per year.}
 \label{fig:Chaos_Tools_Per_Year}
\end{figure*}

\textcolor{blue}{
To contextualize the evolution of these ten tools within the broader CE ecosystem, \Cref{fig:Chaos_Tools_Per_Year} shows the number of distinct \emph{new CE tools introduced per year}, combining the tools in \Cref{tab:chaos_tools_results} with additional CE tools identified from the literature and GitHub. Following the introduction of \textit{Chaos Monkey} in 2010, few new tools emerged until 2015. The number of newly introduced tools increased from 2016 onward, coinciding with the growing adoption of Kubernetes and cloud-native architectures~\citep{CNCF2018,kratzke2017understanding}.}

\textcolor{blue}{
The introduction of new CE tools peaked in 2018, when 21 distinct tools
were introduced. This increase coincided with the growing adoption of
Kubernetes and the expansion of the Cloud Native Computing Foundation
(CNCF) ecosystem~\citep{CNCF2018,kratzke2017understanding}. After 2018,
the number of newly introduced tools declined, while established frameworks
such as \textit{LitmusChaos}, \textit{Chaos Mesh}, and
\textit{Chaos Toolkit} continued to evolve, alongside the increasing
integration of resilience capabilities into cloud-native environments
~\citep{Gremlin2023Tools,IBM2024,owotogbe2025chaos,fogli2023chaos}.
Taken together, this pattern suggests a shift from the introduction of new
CE tools toward the continued development and integration of existing ones.
Similar transitions from tool proliferation toward continued maintenance
have been reported in mutation-testing and data-quality tool ecosystems
~\citep{sanchez2022mutation,ustunboyacioglu2024data}.}
We next examine the evolution of the ten selected CE tools themselves. \Cref{tab:chaos_tools_repo_stats} 
summarizes their repository activity and
version-release histories, providing indicators of development, maintenance, and community activity.

\begin{table*}[ht]
 \footnotesize
 \caption{CE tools under study: FC (first commit), LC (last commit), W (watchers), S (stars), F (forks), C (commits), I (GitHub open-issue count, including open pull requests), total tagged version releases (R), first and latest tagged versions (FR, LR).}
 \label{tab:chaos_tools_repo_stats}
 \centering
 
 \begin{tabular}{lrrrrrrrrrr}
 \hline
 \textbf{Tool Name} & \textbf{FC} & \textbf{LC} & \textbf{W} & \textbf{S} & \textbf{F} & \textbf{C} & \textbf{I} & \textbf{R} & \textbf{FR} & \textbf{LR} \\
 \hline
 
 Chaos Monkey 
 & 2016 
 & 2024 
 & 658 
 & 17,096 
 & 1,304 
 & 165 
 & 34 
 & 3 
 & 2.1.1 
 & 2.1.3 \\
 \hline
 
 Toxiproxy 
 & 2014 
 & 2026 
 & 504 
 & 12,289 
 & 508 
 & 752 
 & 103 
 & 32 
 & 1.0.0 
 & 2.12.0 \\
 \hline
 
 Chaos Mesh 
 & 2019 
 & 2026 
 & 119 
 & 7,862 
 & 1,028 
 & 1,996 
 & 548 
 & 82 
 & 0.8.0 
 & 2.8.4 \\
 \hline
 
 ChaosBlade 
 & 2019 
 & 2026 
 & 176 
 & 6,514 
 & 1,006 
 & 343 
 & 365 
 & 42 
 & 0.0.1 
 & 0.7.0 \\
 \hline
 
 LitmusChaos 
 & 2017 
 & 2026 
 & 83 
 & 5,604 
 & 892 
 & 3,182 
 & 398 
 & 120 
 & 0.6.0 
 & 3.31.0 \\
 \hline
 
 Kube Monkey 
 & 2016 
 & 2026 
 & 53 
 & 3,072 
 & 254 
 & 450 
 & 25 
 & 8 
 & 0.3.0 
 & 0.5.4 \\
 \hline
 
 Pumba 
 & 2016 
 & 2026 
 & 46 
 & 3,141 
 & 215 
 & 713 
 & 3 
 & 90 
 & 0.1.3 
 & 1.2.1 \\
 \hline
 
 Chaos Toolkit 
 & 2017 
 & 2026 
 & 42 
 & 2,021 
 & 206 
 & 452 
 & 3 
 & 58 
 & 0.2.4 
 & 1.20.0 \\
 \hline
 
 PowerfulSeal 
 & 2017 
 & 2021 
 & 47 
 & 1,981 
 & 164 
 & 1,150 
 & 55 
 & 36 
 & 1.0.1 
 & 3.3.0 \\
 \hline
 
 Chaoskube 
 & 2016 
 & 2026 
 & 21 
 & 1,934 
 & 126 
 & 1,008 
 & 34 
 & 51 
 & 0.4.0 
 & 0.39.1 \\
 \hline
 
 \end{tabular}
\end{table*}


The data reveal notable variation in development patterns. LitmusChaos has the most tagged version releases (120), followed by Pumba (90) and Chaos Mesh (82), indicating more frequent release activity than the other studied tools. LitmusChaos and Chaos Mesh have been accepted into the Cloud Native Computing Foundation (CNCF), an open-source governance body that hosts cloud native projects~\citep{CNCF2025,deng2024cloud,jakobczyk2020cloud}. Their current incubating maturity level reflects the increasing adoption, active maintainers, and alignment with CNCF standards~\citep{CNCFLitmus2025,CNCFChaosMesh2025}. ChaosBlade (42 releases) is a CNCF Sandbox project with emerging community support, while Toxiproxy, Pumba, Chaos Toolkit, and Chaoskube exhibit sustained development and regular maintenance typical of independent utilities. In contrast, Chaos Monkey (3 releases since 2016) appears stable and mature, and PowerfulSeal has shown no activity since 2021, suggesting dormancy.

\textcolor{blue}{
\Cref{tab:chaos_tools_repo_stats} provides complementary information on
the development and release histories of the selected CE tools. The first
and last commit years characterize their observed development periods,
whereas the number of tagged releases and the first and latest tagged
versions characterize their release histories. For example,
\textit{Toxiproxy} has an observed commit history from 2014 to 2026 and
32 tagged releases, ranging from version 1.0.0 to 2.12.0. Similarly,
\textit{Pumba} has an observed commit history from 2016 to 2026 and
90 tagged releases, ranging from version 0.1.3 to 1.2.1.
}

Overall, the timeline analysis shows two dominant evolutionary patterns. First, tools under the CNCF umbrella (\textit{LitmusChaos}, \textit{Chaos Mesh}, \textit{ChaosBlade}) exhibit rapid iteration and structured governance, signaling institutionalized development and growing community adoption. Second, independently maintained tools (\textit{Toxiproxy}, \textit{Pumba}, \textit{Chaos Toolkit}) exhibit slower but sustained evolution, emphasizing stability and long-term use over innovation cycles. In contrast, frameworks such as \textit{Chaos Monkey} and \textit{Powerfulseal} 
appear largely stable or inactive, pointing to lifecycle maturity or declining relevance.

\begin{tcolorbox}[colback=blue!3, colframe=black!50, boxrule=0.8pt,
arc=1pt, left=4pt, right=4pt, top=2pt, bottom=2pt]
\textbf{RQ2:}
The CE ecosystem peaked in new-tool introductions in 2018 and has since shifted
toward the continued development of established tools. Across the ten selected
tools, release and repository activity vary substantially, with
\textit{LitmusChaos} and \textit{Chaos Mesh} showing the highest
version-release activity.
\end{tcolorbox}

\subsection{RQ3: What Open-source Projects Adopt CE Tools?}
\label{sec:Repo_Class}

To answer RQ3, we study who uses CE tools and for what purposes, whether adoption is concentrated among industry practitioners, academic researchers, or educational users. Understanding this distinction clarifies the extent to which CE practices have diffused across the broader software ecosystem and whether CE has transitioned from a specialized concern within reliability engineering to a general development concern. Building on RQ2, RQ3 examines the adoption contexts of CE tools by analyzing which projects use them and who maintains them, to determine whether CE extends beyond specialized reliability teams into open-source and educational communities. We manually classified validated repositories by purpose and ownership, based on owners' backgrounds, and analyzed adoption trends across groups. Following~\citet{sanchez2022mutation}, repositories were grouped into \textit{Teaching}, \textit{Learning}, \textit{Research}, \textit{Development}, and \textit{Unclassified}

Repositories were classified as Teaching if they were developed to instruct others, such as course materials, workshop guides, or other content created for classroom or training use\footnote{Teaching repository: \url{https://github.com/michaelhyatt/k8s-o11y-workshop}}. The Learning category included repositories intended for self-guided exploration, such as tool demos, example code, or tutorials\footnote{Learning repository: \url{https://github.com/EleSangwon/ChaosMesh}}. We acknowledge that some repositories include both teaching-oriented and exploratory content. In such cases, classification was guided by the primary stated purpose of the repository, as specified in its README or description. If the goal appeared to be delivering instruction (e.g., as part of a course or workshop), it was classified as Teaching; if the focus was on individual experimentation or tool demonstration, it was classified as Learning. Repositories supporting experimental implementations or academic investigations were categorized under Research\footnote{Research repository: \url{https://github.com/VladCroitoru/dockerfile_smells_project}}.

Development repositories integrate CE tools into software projects or demonstrate their use in development workflows, including repositories focused on related learning activities. \footnote{Development repository: \url{https://github.com/FoundationDB/fdb-kubernetes-operator}}.
Unclassified repositories lacked sufficient information, had ambiguous names, or were written in non-English languages.\footnote{Unclassified repository: \url{https://github.com/xymeng16/study}}. Despite reviewing names, descriptions, documentation, and source code, some repositories could not be reliably categorized, consistent with prior observations~\citep {sanchez2022mutation,ustunboyacioglu2024data}.
In total, we analyzed 1,275 repositories that referenced the 10 CE tools under study.

\begin{center}
\includegraphics[width=0.40\textwidth]{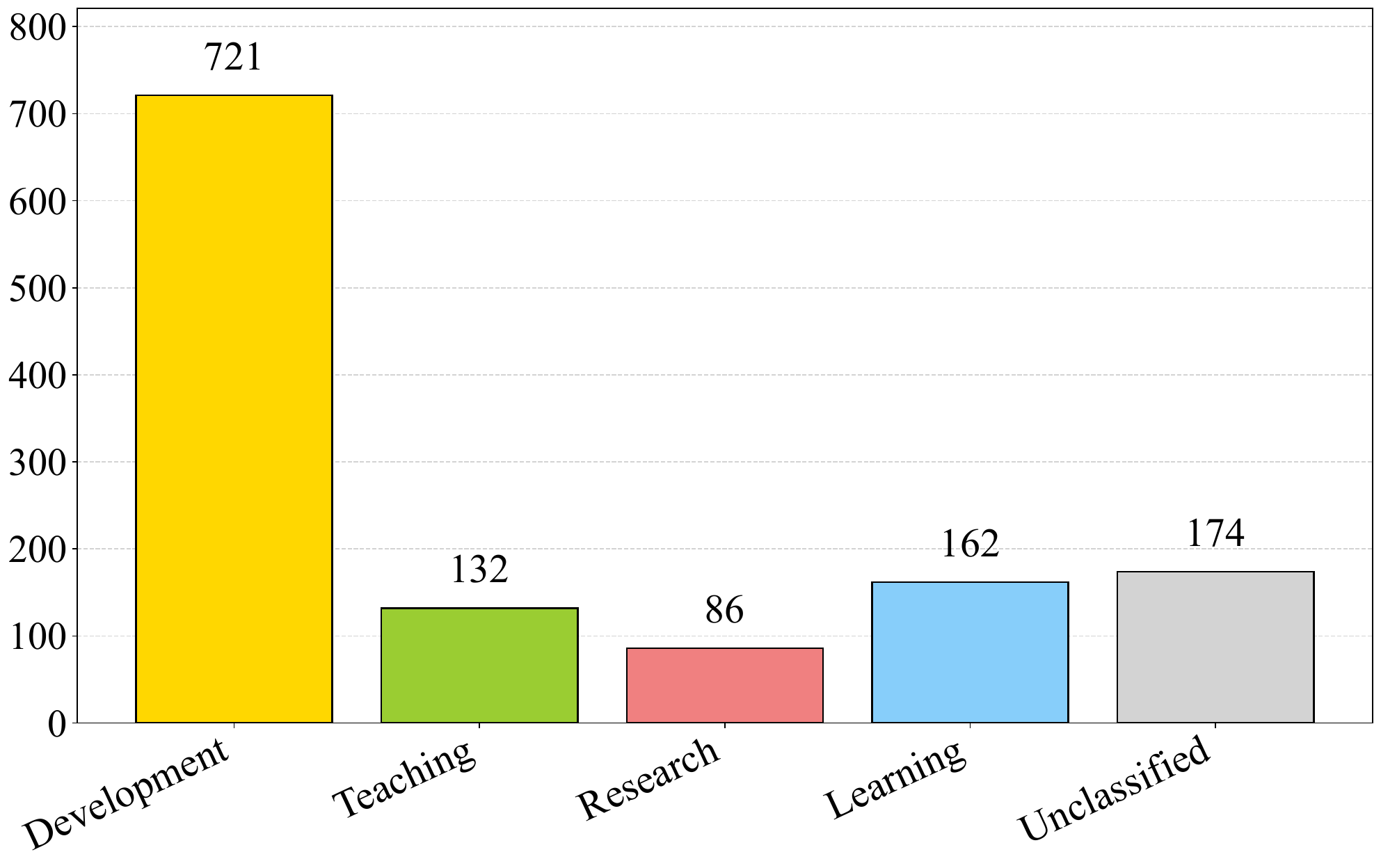}
\captionof{figure}{Classification of repositories by type}
\label{fig:CE_Repo_Type_Distribution}
\end{center}

\Cref{fig:CE_Repo_Type_Distribution} illustrates the distribution of repository types across all CE tools. 
Development repositories are the most common, accounting for 56.55\% (721 of 1,275), followed by Learning with 12.71\% (162), Teaching with 10.35\% (132), and Research with 6.75\% (86). Only 13.65\% (174) remained Unclassified. This distribution shows that CE tools are used predominantly in software development contexts, confirming that most practitioners apply them to build or test resilient systems rather than for educational or exploratory purposes. The small proportion of Research repositories suggests that academic use remains secondary to practical deployment.


\Cref{fig:purpose_origin} (a) shows the breakdown of these categories for each tool. Development is the dominant use for most tools, particularly for Toxiproxy (260 repositories) and Chaos Mesh (176). Teaching usage is highest in Chaos Monkey (35) and Chaos Mesh (32), indicating strong adoption in educational contexts. Learning repositories are present in moderate numbers, led by Chaos Monkey (43) and Chaos Mesh (40). Research usage is minimal, with Chaos Mesh (29) and Chaos Monkey (19) being the most notable. Many tools, including Pumba, Kube-monkey, and Chaoskube, have few or no research-related repositories. The Unclassified category is substantial for tools such as Toxiproxy (44), Chaos Monkey (33), Chaos Mesh (32), and LitmusChaos (27), suggesting limited documentation or undefined usage patterns. Across most tools, Development dominates. Several tools, such as \textit{Chaos Monkey}, \textit{Chaos Mesh}, and \textit{LitmusChaos}, span all repository categories, indicating their broad ecosystem engagement. In contrast, tools such as \textit{PowerfulSeal} and \textit{Chaoskube} show narrower distributions across repository purposes, suggesting more task-specific adoption.

 \begin{figure*}[ht]
    \centering
    \includegraphics[width=0.95\textwidth]{Purpose_origin.pdf}
    \caption{Distribution of CE-adopting repositories across the selected tools by
    (a) purpose and (b) origin.}
    \label{fig:purpose_origin}
\end{figure*}

To provide deeper insights, the repositories were classified by ownership origin using the scheme employed by comparable studies~\citep{sanchez2022mutation,ustunboyacioglu2024data}. Repositories related to academic activities were classified as Academia\footnote{Academic repository:~\url{https://github.com/fglmtt/tiot-2023-artifacts}}, those affiliated with public institutions or open-source communities as Public\footnote{Public repository:~\url{https://github.com/HeeManSu/Meshery}}. Additionally, repositories are classified as Industry repositories\footnote{Industry repository:~\url {https://github.com/porco-rosso-j/aztec-packages-aztec-bb}} if they are associated with a company, or if they represent an open-source project officially maintained or presented by a company, and those owned by individuals or unclassified entities as Personal\footnote{Personal repository: \url{https://github.com/kubevious/mock-data}}. 

\Cref{fig:purpose_origin} (b) classifies repositories by owner origin: Academia, Industry, Personal, Public, and Unclassified. Industry-affiliated repositories are prominent across the tool landscape, including Toxiproxy (177), Chaos Mesh (111), and ChaosBlade (22). Personal ownership is also common, particularly for Chaos Monkey (97), Chaos Mesh (87), and LitmusChaos (63). Academic contributions are limited, with Chaos Monkey (36), Chaos Mesh (29), and Toxiproxy (21) leading, whereas tools such as Chaos Toolkit and Chaoskube show little academic presence. Public repositories are most prominent for Chaos Toolkit (32) and Chaoskube (24), indicating engagement from open-source communities and public institutions. The tool-level breakdown highlights the dominance of industrial actors, particularly around Kubernetes-native and microservice-level tools such as Chaos Mesh and Toxiproxy. Widespread individual contributions across multiple tools indicate an active developer community extending CE utilities beyond formal organizations. Public participation concentrates in tools with open governance, such as Chaos Toolkit, while academic involvement remains limited and tool-specific, underscoring the need for closer industry–academia collaboration to formalize CE practices and evaluation.

\Cref{fig:CE_Repo_Origin_Distributio} shows the ownership distribution of CE repositories. Industry accounts for 34.82\% (444/1,275), followed by personal repositories at 33.49\% (427), public repositories at 14.67\% (187), academia at 8.16\% (104), and Unclassified repositories at 8.86\% (113). This distribution indicates that CE development and adoption are driven mainly by industry and individual practitioners, with limited academic involvement, suggesting that CE has evolved primarily as an operational discipline rather than a research-focused field~\citep{garousi2019characterizing,marijan2022industry}.


\begin{center}
\includegraphics[width=0.40\textwidth]{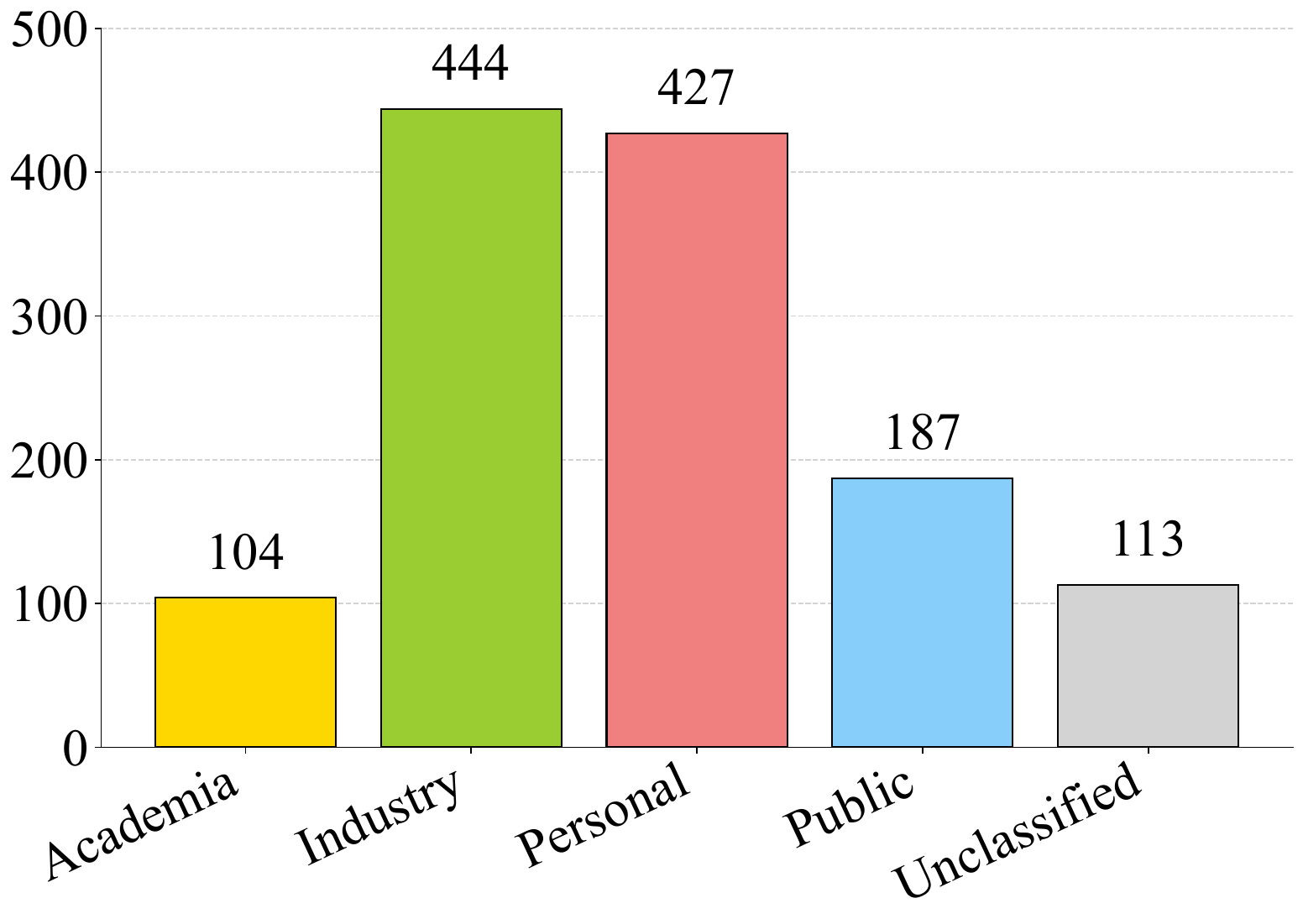}
\captionof{figure}{Classification of repositories by origin}
\label{fig:CE_Repo_Origin_Distributio}
\end{center}

\begin{tcolorbox}[colback=blue!3, colframe=black!50, boxrule=0.8pt, arc=1pt, left=4pt, right=4pt, top=2pt, bottom=2pt]
\textbf{RQ3:} Most CE repositories support development, emphasizing resilience testing in existing systems. Although tools such as Chaos Monkey and LitmusChaos appear across categories, academic and public-sector use remains limited, exposing a research–practice gap and an opportunity to strengthen CE through education and research alongside industry-led adoption.
\end{tcolorbox}

\subsection{RQ4: How Active Are the Open-source Projects that Adopt CE Tools, and What Evidence Reflects their Prominence in the Developer Community?}

To address RQ4, we examine the activity and relevance of repositories that employ CE tools, using established GitHub metrics to assess sustained development, maintenance, and popularity across the open-source ecosystem. We aim to determine whether CE tool adoption represents long-term engineering integration or short-lived experimentation. To address this question, we analyze repository activity for the selected CEtools using standard GitHub indicators, including commits, contributors, watchers, stars, and forks, following~\citet{sanchez2022mutation}. These metrics are used to characterize repository activity and community engagement. Commit-count data were available for 1,231 of the 1,275 validated
repository-tool records. Among these, 24.9\% had at most 10 commits and
51.4\% had at most 50 commits, whereas 14.5\% had more than 1,000 commits.
These results show substantial variation in development activity across the
analyzed repositories.


\begin{figure*}[ht]
    \centering
    \includegraphics[width=\textwidth]{commit_tools_activity.pdf}
    \caption{Commit-count distribution across CE-adopting repositories by
    (a) CE tool and (b) repository purpose. Stacked bars show the number of
    repositories within each commit-count range.}
    \label{fig:Commit_distribution_t}
\end{figure*}

\textcolor{blue}{
\Cref{fig:Commit_distribution_t} (a) shows substantial variation in activity
across CE-adopting repositories. Toxiproxy and Chaos Mesh have the
strongest representation among highly active repositories, whereas
repositories adopting several other tools are concentrated in lower
commit ranges. Overall, CE adoption spans both short-lived projects and
repositories with sustained development activity.
}

\textcolor{blue}{
This concentration of sustained activity around a limited set of tools
is consistent with patterns reported in other open-source ecosystems,
where a few mature projects account for substantial development
activity~\citep{kuwata2014study,zhou2016inflow,coelho2017modern}.
}

\textcolor{blue}{
\Cref{fig:Commit_distribution_t} (b) shows that repository activity differs
by project purpose. Development repositories are more strongly represented
in the higher commit ranges, whereas teaching, learning, and unclassified
repositories are concentrated in the lower ranges. Research repositories
show a more distributed activity profile. These patterns suggest that
sustained development activity is more characteristic of development-oriented
CE-adopting repositories than of educational or exploratory ones.
}

\begin{figure*}[ht]
 \centering
 \begin{subfigure}{0.30\linewidth} 
 \centering
 \includegraphics[width=\linewidth]{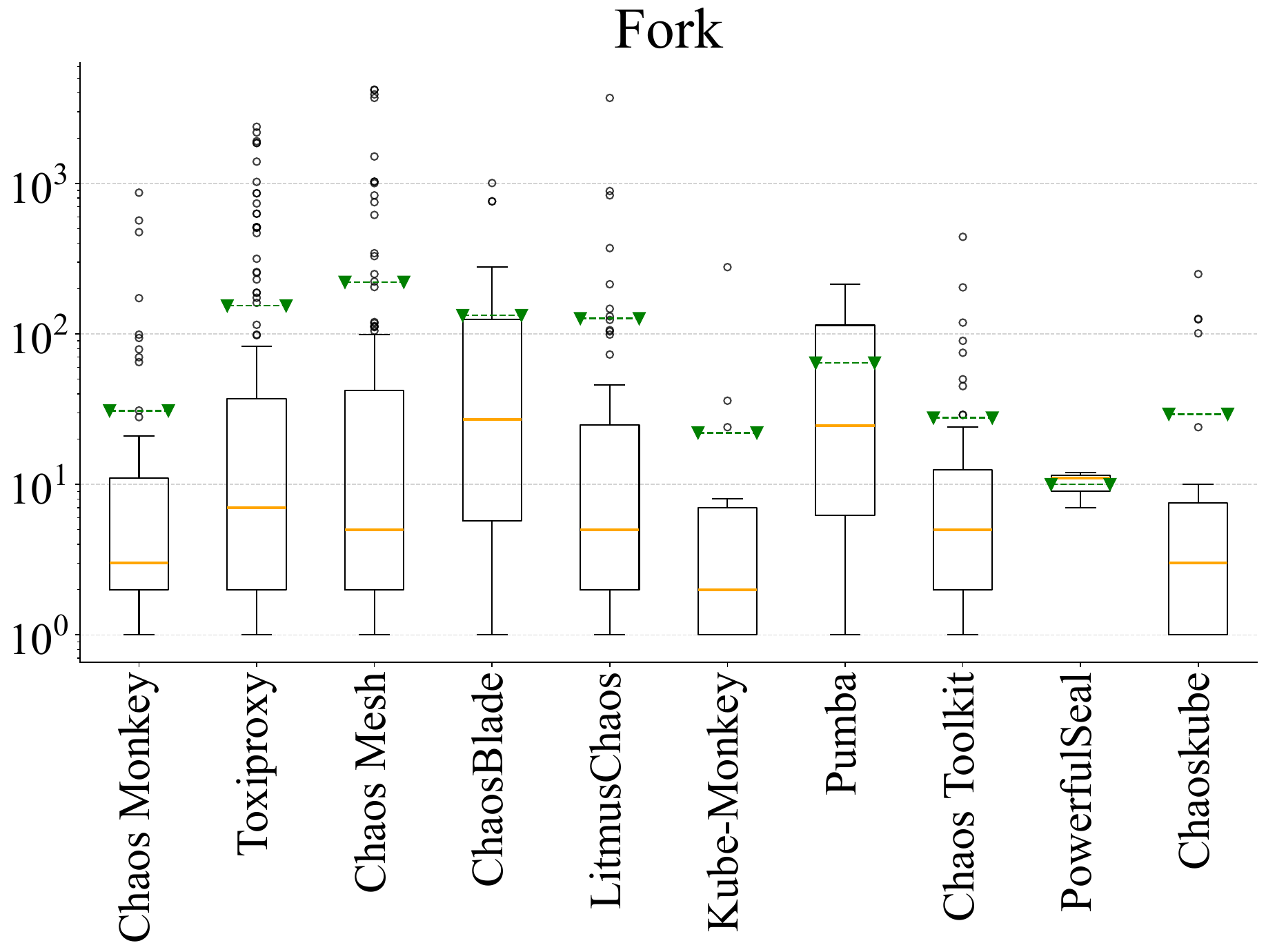} %
 \end{subfigure}
 \begin{subfigure}{0.30\linewidth}
 \centering
 \includegraphics[width=\linewidth]{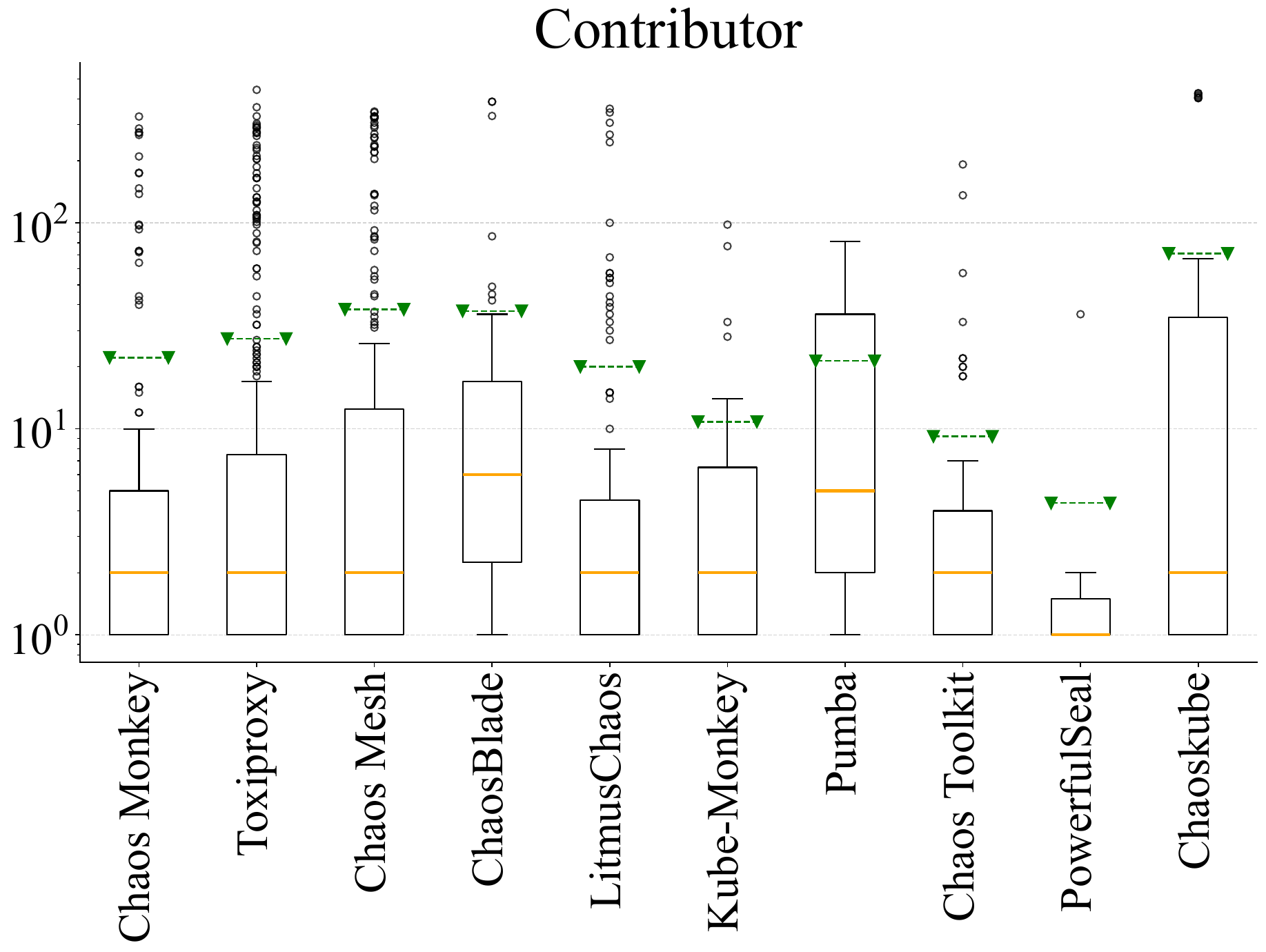} %
 \end{subfigure}
 
 \begin{subfigure}{0.30\linewidth}
 \centering
 \includegraphics[width=\linewidth]{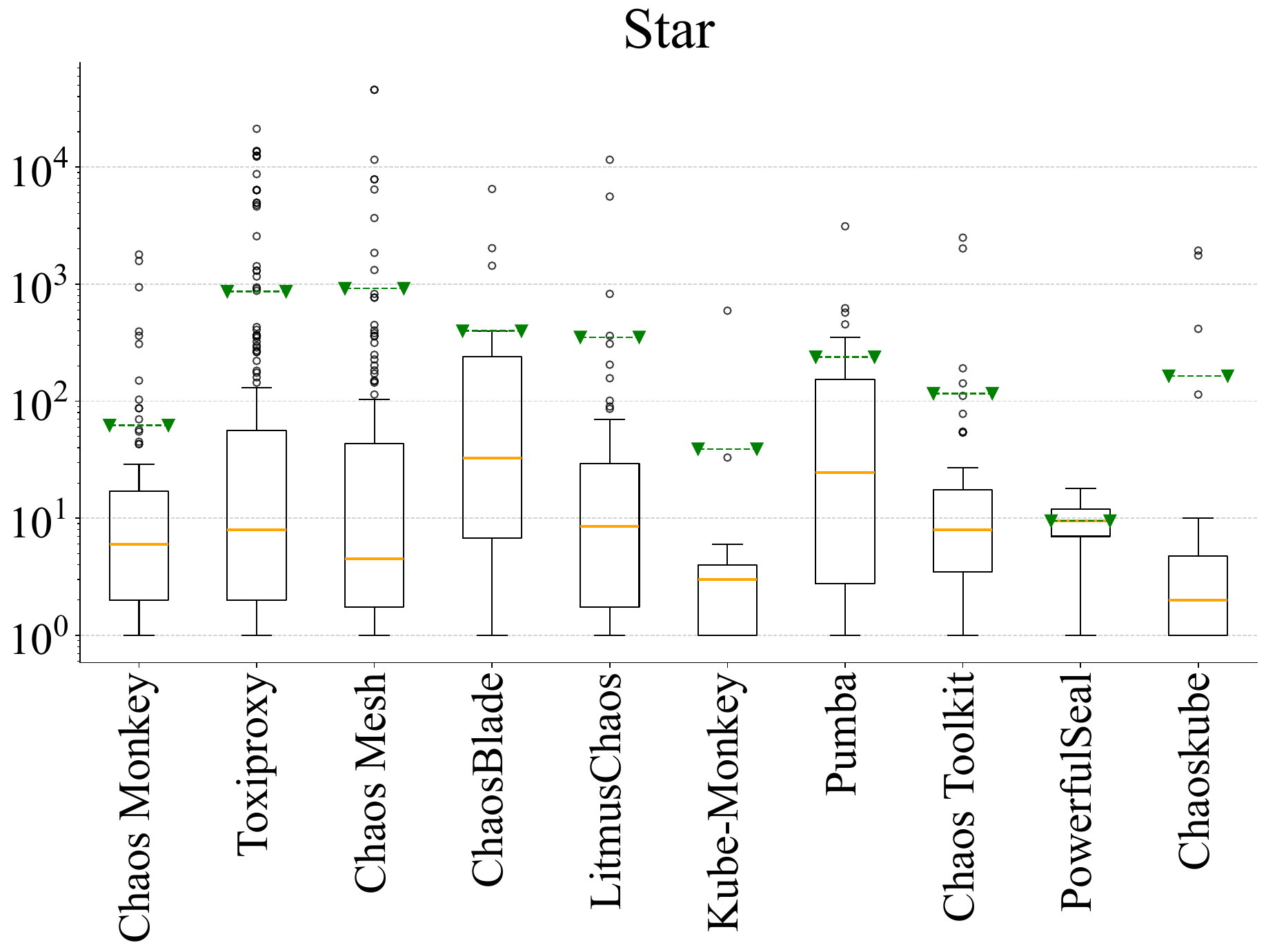} %
 \end{subfigure}
 \begin{subfigure}{0.30\linewidth}
 \centering
 \includegraphics[width=\linewidth]{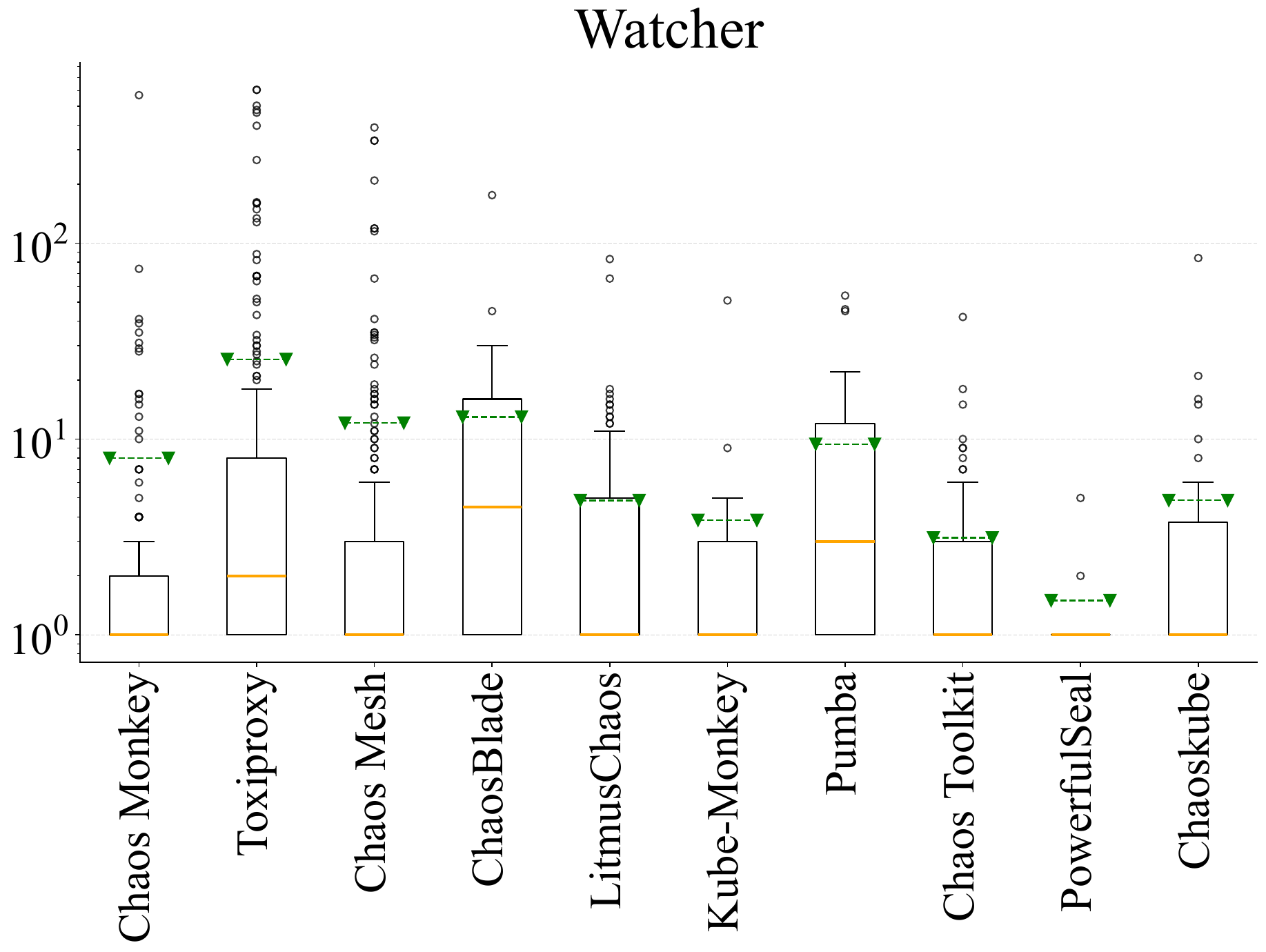} %
 \end{subfigure}

 \caption{Distribution of forks,  contributors, stars, and watchers in the selected tools.}
 \label{fig:Commit_distribution_metrics}
\end{figure*}


\Cref{fig:Commit_distribution_metrics} illustrates the mean, median, minimum, maximum, and outlier values for contributors, watchers, stars, and forks in the repositories using the selected CE tools. The Y-axis is logarithmically scaled to improve visualization across a wide range of values. Most distributions exhibit low positive values, typically between 1 and 10, with significant outliers indicating highly active or popular projects. Projects with more than 50 contributors, watchers, stars, or forks were classified as highly popular, following the approach of~\citet{sanchez2022mutation}. 

The analysis shows that most repositories have relatively few contributors, as reflected by low median values across the tools. Nevertheless, substantial outliers occur, with maximum contributor counts reaching 442 for Toxiproxy, 426 for Chaoskube, 387 for ChaosBlade, 358 for LitmusChaos, and 347 for Chaos Mesh. Watcher activity is similarly concentrated at low values, with most tools having median positive values between one and five. However, notable outliers are observed for Toxiproxy (607 watchers), Chaos Monkey (570), and Chaos Mesh (390), indicating that a small subset of repositories attracts substantially greater passive engagement. In contrast, stars and forks provide stronger indicators of popularity and reuse. Chaos Mesh reaches the highest observed values, with up to 45,740 stars and 4,193 forks, indicating extensive visibility and downstream reuse. Toxiproxy also shows substantial engagement, reaching 21,238 stars and 2,383 forks, while LitmusChaos reaches 11,550 stars and 3,705 forks. ChaosBlade also exhibits notable engagement, with up to 6,497 stars and 1,007 forks. These indicators highlight their prominence within the open-source CE community. In contrast, tools such as Kube-Monkey and PowerfulSeal maintain lower visibility, with substantially smaller maximum star and fork counts.

These popularity metrics confirm the dominance of a few high-visibility frameworks within the CE landscape. The concentration of stars and forks around Chaos Mesh, Toxiproxy, and LitmusChaos shows convergence toward mature, community-supported ecosystems. Conversely, most other tools attract limited engagement, suggesting that CE awareness has grown faster than broad, active participation.

\begin{tcolorbox}[colback=blue!3, colframe=black!50, boxrule=0.8pt, arc=1pt, left=4pt, right=4pt, top=2pt, bottom=2pt]
\textbf{RQ4:} 
Repository activity and popularity vary substantially across CE-adopting repositories. A smaller set of repositories, particularly those adopting Toxiproxy and Chaos Mesh, exhibits sustained development activity and greater community prominence, whereas many repositories show comparatively limited activity and engagement.
\end{tcolorbox}


\subsection{RQ5: Which Capabilities of CE Tools Are Exploited by Open-source Projects?}

To address RQ5, we examine how CE tool capabilities are actually exercised in real-world projects, particularly in open-source settings, where usage patterns can reveal which fault types developers consider most relevant. 
Understanding this distribution helps determine whether CE is primarily practiced at the infrastructure level (e.g., network or instance faults) or has evolved toward higher-level, application-centric resilience testing. 
Our findings provide empirical evidence on which tool capabilities developers most often exploit in practice.

\begin{table*}[ht]
 \caption{Fault types and descriptions. Instance termination (Inst. termination), network faults (Net. faults), resource faults (Res. faults), application fault (App. faults), TCP (transmission control protocol), HTTP (hypertext transfer protocol), DNS (domain name system), IP (internet protocol).}
 \label{tab:fault_types}
 \tiny
 \begin{tabular}{|p{0.18\textwidth}|p{0.23\textwidth}|p{0.47\textwidth}|}
 \hline
 \textbf{Category} & \textbf{Fault type} & \textbf{Description} \\
 \hline
 Network Faults & Packet loss & Drops data packets during transmission. \\
 \cline{2-3}
 & Network delay & adds latency to network communications. \\
 \cline{2-3}
 & Bandwidth limitation & Caps the speed or data transfer rate. \\
 \cline{2-3}
 & Packet duplication & Creates duplicates of transmitted packets. \\
 \cline{2-3}
 & Network partition & Splits the network into isolated segments. \\
 \cline{2-3}
 & Network down & Disables a service or network endpoint. \\
 \cline{2-3}
 & Network latency & Introduces a delay in network traffic. \\
 \cline{2-3}
 & Slow close & Delays closing the connection after data transmission. \\
 \cline{2-3}
 & Network timeout & Stops data transmission and optionally closes the connection after a timeout. \\
 \cline{2-3}
 & Network reset peer & Simulates an abrupt TCP connection reset. \\
 \cline{2-3}
 & Network slicer & Splits packets into smaller fragments with optional delays between them. \\
 \cline{2-3}
 & HTTP fault & Introduces HTTP-layer errors or behaviors (requires custom logic or upstream modification). \\
 \cline{2-3}
 & Packet corruption & Randomly corrupts packet data. \\
 \cline{2-3}
 & Request abort & Stops a request midway. \\
 \cline{2-3}
 & Packet reordering & Changes the order of packets being delivered. \\
 \cline{2-3}
 & DNS fault & Simulates DNS resolution failure or maps domains to incorrect IPs. \\
 \cline{2-3}
 & Network jitter & Adds random changes to network delay — some packets arrive fast, some slow, like a very unstable connection. \\
 \cline{2-3}

 \hline
 Instance Termination Faults & Stopping & Pausing a running process without deletion. \\
 \cline{2-3}
 & Termination & Permanently shuts down and deletes a component. \\
 \cline{2-3}
 & Restart & Stops and immediately restarts a component. \\
 \cline{2-3}
 & Draining & Clears active tasks and blocks new ones. \\
 \cline{2-3}
 & Detachment & Disconnects a component from its dependencies. \\
 \hline
 Resource Stress Faults & Disk loss & Disconnects a disk from the system. \\
 \cline{2-3}
 & CPU stress & Overloads the CPU with intensive tasks. \\
 \cline{2-3}
 & Memory stress & Consumes system memory to simulate high usage. \\
 \cline{2-3}
 & Disk storage fill & Fills storage with random data to limit free space. \\
 \hline
 Application-level Faults & Method call delay & Delays the start of a method call. \\
 \cline{2-3}
 & Process stop & Forcefully stops the running application process. \\
 \cline{2-3}
 & Return value modification & Alter method return values to simulate incorrect outputs. \\
 \cline{2-3}
 & Exception & Simulates an exception during execution. \\
 \hline
 \end{tabular}
\end{table*}

We manually analyzed 2,410 fault-injection instances across 1,275
CE-adopting repositories associated with the selected tools. Based on the
official tool documentation, we identified the supported fault types,
parameters, and scope, and classified each observed fault instance according
to the taxonomy in \Cref{tab:fault_types}. First, \textit{network faults}
introduce disruptions such as latency, packet loss, and connection
interruptions~\citep{sondhi2021chaos,jernberg2020getting,yu2024survey}.
Supported by tools such as Toxiproxy and Chaos Mesh, they are the most common
category, accounting for 44.85\% (1,081/2,410) of observed fault instances.
Second, \textit{instance termination faults} remove components such as
Kubernetes pods, nodes, or cloud instances, exposing service unavailability,
failover, recovery, and rescheduling behavior
~\citep{ribeiro2024chaos,jernberg2020getting}. Tools such as Kube-monkey and
Chaoskube support these faults, which account for 29.96\% (722/2,410).
Third, \textit{resource stress faults} assess system behavior under constrained
CPU, memory, or related resources
~\citep{lewis2019chaos,basiri2016chaos,jernberg2020getting}. Supported by
tools such as ChaosBlade and Pumba, they account for 22.61\% (545/2,410).
Fourth, \textit{application-level faults} target application logic and error
handling, including delayed execution and exception-related failures
~\citep{jernberg2020getting,chaos_engineering_chapter8,pawlikowski2025chaos}.
Supported by tools such as Chaos Mesh, PowerfulSeal, and Pumba, they are the
least common category, accounting for 2.57\% (62/2,410).

\Cref{tab:Fault_Categories} summarizes how types of faults are used in the selected repositories. Mapping tools to fault categories shows that Toxiproxy and Chaos Mesh dominate network faults; Chaos Monkey, LitmusChaos, and Chaos Mesh account for much of instance termination; LitmusChaos, Chaos Mesh, and ChaosBlade account for most resource stress; and ChaosBlade and Chaos Mesh account for most application-level faults. Overall, chaos engineering remains dominated by infrastructure-focused tools, consistent with what we reported in RQ1 to RQ4.

\definecolor{lightgray}{gray}{0.8}

\begin{table*}[ht]
\centering
\caption{Usage (count) of fault types supported by the selected CE tools in the GitHub repositories. Gray cells indicate fault types not supported by the corresponding tool. Zeros in non-gray cells indicate fault types supported but not observed in the analyzed repositories.}
\renewcommand{\arraystretch}{1.2}

\begin{adjustbox}{width=\linewidth, center}

\begin{tabular}{|l|c|c|c|c|c|c|c|c|c|c|c|c|c|c|c|c|c|c|c|c|c|c|c|c|c|c|c|}

\hline

& \multicolumn{3}{|c|}{\textbf{Instance termination}}
& \multicolumn{16}{|c|}{\textbf{Network faults}}
& \multicolumn{3}{|c|}{\textbf{Resource faults}}
& \multicolumn{4}{|c|}{\textbf{Application faults}}
& \\ \hline

\textbf{Tool}

& \begin{turn}{90}Container fault\end{turn}
& \begin{turn}{90}Pod termination\end{turn}
& \begin{turn}{90}Virtual machine\end{turn}

& \begin{turn}{90}Network delay\end{turn}
& \begin{turn}{90}Network partition\end{turn}
& \begin{turn}{90}Network corruption\end{turn}
& \begin{turn}{90}Network loss\end{turn}
& \begin{turn}{90}Bandwidth control\end{turn}
& \begin{turn}{90}Network duplicate\end{turn}
& \begin{turn}{90}Request abort\end{turn}
& \begin{turn}{90}Slow close\end{turn}
& \begin{turn}{90}Timeout\end{turn}
& \begin{turn}{90}Reset peer\end{turn}
& \begin{turn}{90}Network reordering\end{turn}
& \begin{turn}{90}Slicer\end{turn}
& \begin{turn}{90}Limit data\end{turn}
& \begin{turn}{90}Down\end{turn}
& \begin{turn}{90}DNS fault\end{turn}
& \begin{turn}{90}HTTP fault\end{turn}

& \begin{turn}{90}CPU stressor\end{turn}
& \begin{turn}{90}Memory stressor\end{turn}
& \begin{turn}{90}Disk fault\end{turn}

& \begin{turn}{90}Exception injection\end{turn}
& \begin{turn}{90}Return value modification\end{turn}
& \begin{turn}{90}Process stop\end{turn}
& \begin{turn}{90}Method call delay\end{turn}

& \begin{turn}{90}\textbf{Total}\end{turn}
\\ \hline


Chaos Monkey

& 0
& \cellcolor{lightgray}
& 189

& \cellcolor{lightgray}
& \cellcolor{lightgray}
& \cellcolor{lightgray}
& \cellcolor{lightgray}
& \cellcolor{lightgray}
& \cellcolor{lightgray}
& \cellcolor{lightgray}
& \cellcolor{lightgray}
& \cellcolor{lightgray}
& \cellcolor{lightgray}
& \cellcolor{lightgray}
& \cellcolor{lightgray}
& \cellcolor{lightgray}
& \cellcolor{lightgray}
& \cellcolor{lightgray}
& \cellcolor{lightgray}

& \cellcolor{lightgray}
& \cellcolor{lightgray}
& \cellcolor{lightgray}

& \cellcolor{lightgray}
& \cellcolor{lightgray}
& \cellcolor{lightgray}
& \cellcolor{lightgray}

& 189
\\ \hline


Toxiproxy

& \cellcolor{lightgray}
& \cellcolor{lightgray}
& \cellcolor{lightgray}

& 210
& 0
& \cellcolor{lightgray}
& \cellcolor{lightgray}
& 60
& \cellcolor{lightgray}
& 0
& 14
& 103
& 10
& \cellcolor{lightgray}
& 13
& 24
& 1
& \cellcolor{lightgray}
& \cellcolor{lightgray}

& \cellcolor{lightgray}
& \cellcolor{lightgray}
& \cellcolor{lightgray}

& \cellcolor{lightgray}
& \cellcolor{lightgray}
& \cellcolor{lightgray}
& \cellcolor{lightgray}

& 435
\\ \hline


Chaos Mesh

& 14
& 84
& 21

& 119
& 66
& 22
& 63
& 41
& 11
& 4
& 0
& 0
& 0
& 0
& \cellcolor{lightgray}
& \cellcolor{lightgray}
& \cellcolor{lightgray}
& 29
& 28

& 65
& 64
& 32

& 6
& 6
& \cellcolor{lightgray}
& \cellcolor{lightgray}

& 675
\\ \hline


ChaosBlade

& 0
& 10
& \cellcolor{lightgray}

& 19
& 0
& 5
& 16
& 0
& 5
& \cellcolor{lightgray}
& \cellcolor{lightgray}
& \cellcolor{lightgray}
& \cellcolor{lightgray}
& 5
& \cellcolor{lightgray}
& \cellcolor{lightgray}
& \cellcolor{lightgray}
& \cellcolor{lightgray}
& \cellcolor{lightgray}

& 26
& 22
& 41

& 4
& \cellcolor{lightgray}
& 23
& 5

& 181
\\ \hline


LitmusChaos

& 45
& 55
& 96

& 26
& 1
& 18
& 23
& \cellcolor{lightgray}
& 11
& \cellcolor{lightgray}
& \cellcolor{lightgray}
& \cellcolor{lightgray}
& \cellcolor{lightgray}
& \cellcolor{lightgray}
& \cellcolor{lightgray}
& \cellcolor{lightgray}
& \cellcolor{lightgray}
& 11
& \cellcolor{lightgray}

& 73
& 48
& 85

& 0
& \cellcolor{lightgray}
& 0
& \cellcolor{lightgray}

& 492
\\ \hline


Kube-Monkey

& \cellcolor{lightgray}
& 26
& \cellcolor{lightgray}

& \cellcolor{lightgray}
& \cellcolor{lightgray}
& \cellcolor{lightgray}
& \cellcolor{lightgray}
& \cellcolor{lightgray}
& \cellcolor{lightgray}
& \cellcolor{lightgray}
& \cellcolor{lightgray}
& \cellcolor{lightgray}
& \cellcolor{lightgray}
& \cellcolor{lightgray}
& \cellcolor{lightgray}
& \cellcolor{lightgray}
& \cellcolor{lightgray}
& \cellcolor{lightgray}
& \cellcolor{lightgray}

& \cellcolor{lightgray}
& \cellcolor{lightgray}
& \cellcolor{lightgray}

& \cellcolor{lightgray}
& \cellcolor{lightgray}
& \cellcolor{lightgray}
& \cellcolor{lightgray}

& 26
\\ \hline


Pumba

& 44
& 4
& 1

& 0
& 10
& 0
& 10
& \cellcolor{lightgray}
& 2
& 0
& \cellcolor{lightgray}
& 0
& \cellcolor{lightgray}
& 0
& \cellcolor{lightgray}
& 0
& \cellcolor{lightgray}
& \cellcolor{lightgray}
& \cellcolor{lightgray}

& 4
& 4
& 16

& \cellcolor{lightgray}
& \cellcolor{lightgray}
& 12
& \cellcolor{lightgray}

& 108
\\ \hline


Chaos Toolkit

& 6
& 20
& 15

& 13
& 1
& 3
& 5
& 1
& 0
& 0
& 0
& 1
& 1
& 0
& 0
& 0
& 0
& 2
& 3

& 10
& 7
& 4

& 1
& \cellcolor{lightgray}
& 1
& \cellcolor{lightgray}

& 94
\\ \hline


PowerfulSeal

& 5
& 13
& 26

& 22
& 2
& 7
& 22
& \cellcolor{lightgray}
& 6
& \cellcolor{lightgray}
& \cellcolor{lightgray}
& \cellcolor{lightgray}
& \cellcolor{lightgray}
& \cellcolor{lightgray}
& \cellcolor{lightgray}
& \cellcolor{lightgray}
& 0
& 11
& \cellcolor{lightgray}

& 9
& 11
& 24

& \cellcolor{lightgray}
& \cellcolor{lightgray}
& 4
& \cellcolor{lightgray}

& 162
\\ \hline


Chaoskube

& \cellcolor{lightgray}
& 48
& \cellcolor{lightgray}

& \cellcolor{lightgray}
& \cellcolor{lightgray}
& \cellcolor{lightgray}
& \cellcolor{lightgray}
& \cellcolor{lightgray}
& \cellcolor{lightgray}
& \cellcolor{lightgray}
& \cellcolor{lightgray}
& \cellcolor{lightgray}
& \cellcolor{lightgray}
& \cellcolor{lightgray}
& \cellcolor{lightgray}
& \cellcolor{lightgray}
& \cellcolor{lightgray}
& \cellcolor{lightgray}
& \cellcolor{lightgray}

& \cellcolor{lightgray}
& \cellcolor{lightgray}
& \cellcolor{lightgray}

& \cellcolor{lightgray}
& \cellcolor{lightgray}
& \cellcolor{lightgray}
& \cellcolor{lightgray}

& 48
\\ \hline


\textbf{Total}

& 114
& 260
& 348

& 409
& 80
& 55
& 139
& 103
& 35
& 4
& 14
& 104
& 11
& 5
& 13
& 24
& 1
& 53
& 31

& 187
& 156
& 202

& 11
& 6
& 40
& 5

& \textbf{2410}
\\ \hline


\textbf{Category Total}

& \multicolumn{3}{c|}{\textbf{722}}

& \multicolumn{16}{c|}{\textbf{1081}}

& \multicolumn{3}{c|}{\textbf{545}}

& \multicolumn{4}{c|}{\textbf{62}}

& \textbf{2410}
\\ \hline

\end{tabular}

\end{adjustbox}

\label{tab:Fault_Categories}

\end{table*}

\begin{tcolorbox}[colback=blue!3, colframe=black!50, boxrule=0.8pt, arc=1pt, left=4pt, right=4pt, top=2pt, bottom=2pt]
\textbf{RQ5:} 
Chaos Engineering projects mainly target network faults and instance termination (74.81\%), followed by resource stress (22.61\%), with few application-level faults (2.57\%). Consistent with RQ1–RQ4, this reflects the focus of widely used tools on infrastructure disruptions, indicating that CE prioritizes infrastructure-level over application-level resilience.
\end{tcolorbox}

\section{Discussion and practical implications}
\label{sec:discussion}

\textcolor{blue}{
The findings offer several observations, reflections, and implications for both research and practice. In this section, we discuss the key insights emerging from our study.}

\subsection{Implications for Practitioners}

Our findings provide practical guidance for selecting CE tools, interpreting
evidence from public repositories, and designing fault-injection strategies.

\textbf{RQ1} shows that CE-tool use in the studied GitHub repositories is
concentrated around a few frameworks, with \textit{Toxiproxy},
\textit{Chaos Mesh}, and \textit{Chaos Monkey} accounting for 68.86\% of
the validated records. We also found that repository creation associated
with CE-tool use became more prominent after 2018. However, repository
frequency alone does not explain why developers select particular tools.
Prior studies suggest that compatibility with existing automation, CI/CD,
and deployment environments can facilitate tool adoption and continued
use~\citep{sanchez2022mutation,ustunboyacioglu2024data,xu2019configuration,simonsson2021observability}.

Complementing this finding, \textbf{RQ2} shows that although the introduction
of new CE tools peaked in 2018 and subsequently declined, established tools
continued to evolve. Among the studied tools, \textit{LitmusChaos},
\textit{Pumba}, and \textit{Chaos Mesh} recorded the highest numbers of
tagged releases. Prior OSS research associates sustained maintenance and
contributor participation with project sustainability
~\citep{jensen2007role,crowston2008free}, while organizational stewardship,
such as CNCF governance, can provide additional support for long-term project
development~\citep{sikha2021oss}. Taken together, RQ1 and RQ2 suggest that
tool prevalence and technical suitability should be interpreted alongside
evidence of continued development and support.

\begin{tcolorbox}[colback=blue!3, colframe=black!50,
boxrule=0.8pt, arc=1pt, left=4pt, right=4pt, top=2pt, bottom=2pt]

\faLaptop\ \textbf{Implication for Practitioners \#1.}

CE-tool selection should combine technical fit with evidence of continued
support. Practitioners should first assess whether a tool supports the target
environment and required fault types, and then consider its maintenance
history, release activity, contributor participation, and governance where
applicable. Repository popularity alone should not be treated as sufficient
evidence that a tool is suitable for a particular system.

\end{tcolorbox}

\textbf{RQ3} shows that CE use in the studied repositories is predominantly
development-oriented. Repository ownership is distributed mainly between
industry (34.82\%) and personal repositories (33.49\%), while academia
accounts for a smaller share (8.16\%). Some repositories could not be
reliably classified because their purpose or ownership was insufficiently
documented. Clear repository documentation can support transparency,
reproducibility, and knowledge transfer
~\citep{wattanakriengkrai2022github,kang2023papers}; nevertheless,
documentation alone does not establish whether a CE experiment was executed
successfully or improved system resilience.

Similarly, \textbf{RQ4} shows substantial variation in the development
activity of repositories using CE. While 51.42\% have at most 50 commits,
14.46\% exceed 1,000 commits. Prior OSS research associates sustained
development activity and contributor participation with project continuity
~\citep{crowston2008free,coelho2017modern,zhou2016inflow}. These indicators
therefore provide useful context about the repositories in which CE appears.
However, repository activity, contributors, stars, forks, and watchers
characterize the project and its community rather than directly measuring the
effectiveness of its CE experiments. Consequently, RQ3 and RQ4 together
highlight the need to distinguish evidence that CE is present in a repository
from evidence that the corresponding practice is operationally effective.

\begin{tcolorbox}[colback=blue!3, colframe=black!50,
boxrule=0.8pt, arc=1pt, left=4pt, right=4pt, top=2pt, bottom=2pt]

\faLaptop\ \textbf{Implication for Practitioners \#2.}

Practitioners reusing CE practices from public repositories should distinguish
between evidence that CE is configured or integrated and evidence that the
corresponding experiments are operationally effective. Documentation,
repository activity, and community-engagement indicators can provide useful
context, but they should not substitute for system-specific validation of the
experiment and its outcomes.

\end{tcolorbox}

\textbf{RQ5} shows that the observed CE practices are strongly concentrated
on infrastructure-level faults. Network faults and instance termination
account for 74.81\% of the observed fault instances, whereas
application-level faults account for only 2.57\%. This imbalance is important
because software and configuration faults are also significant sources of
system failures~\citep{yuan2014simple,gunawi2014bugs}. Consequently,
extensive use of network and infrastructure faults should not automatically
be interpreted as broad failure coverage. A CE campaign may exercise many
infrastructure scenarios while leaving relevant application-level failure
modes unexplored.

\begin{tcolorbox}[colback=blue!3, colframe=black!50,
boxrule=0.8pt, arc=1pt, left=4pt, right=4pt, top=2pt, bottom=2pt]

\faLaptop\ \textbf{Implication for Practitioners \#3.}

Practitioners should design CE fault portfolios according to the failure risks
of the system under test rather than relying predominantly on infrastructure
faults. Where relevant, network, resource, and instance-level experiments
should be complemented with application-level failures so that the selected
fault scenarios cover a broader range of system-specific risks.

\end{tcolorbox}

These implications are derived from the repository evidence examined in this
study and from prior empirical literature; they have not been independently
validated through practitioner interviews, surveys, or workshops. They should
therefore be interpreted as evidence-based implications of the observed
repository patterns rather than as practitioner-validated requirements.

\subsection{Implications for Researchers}

Our findings also identify research opportunities that follow directly from
the observed adoption patterns, repository characteristics, and
fault-injection practices.

\textbf{RQ1} and \textbf{RQ2} reveal two related patterns: repository-visible
CE use is concentrated around a limited number of tools, while established
tools differ in their continued development and release activity. Repository
mining can establish these patterns but cannot explain the mechanisms that
produce them. Prior studies identify organizational support, governance,
coordination, contributor participation, institutional support, and ecosystem
relationships as factors associated with software technology adoption and
project persistence
~\citep{bird2011don,di2025investigating,offerman2022study,santos2020software,
Tamburri2013Organizational,Feng2023State,Yin2022Open}. Future research should
therefore examine how these technical and socio-organizational factors relate
specifically to CE-tool adoption and long-term evolution.

\begin{tcolorbox}[colback=blue!3, colframe=black!50,
boxrule=0.8pt, arc=1pt, left=4pt, right=4pt, top=2pt, bottom=2pt]

\faFlask\ \textbf{Implication for Researchers \#1.}

Researchers should investigate why repository-visible CE use concentrates
around particular tools and which technical, organizational, governance, and
ecosystem factors contribute to their continued development and adoption.

\end{tcolorbox}

\textbf{RQ3} shows that academic-owned repositories constitute a
comparatively small share of the studied dataset. This observation should not
be interpreted as evidence that academia participates little in CE overall,
because repository ownership captures only one observable aspect of
participation. Nevertheless, the pattern motivates further investigation into
why academic-owned CE repositories are less represented and how CE knowledge
and tools move between research and practice. Prior research identifies
differences in incentives and research-practice collaboration as relevant
issues in empirical software engineering
~\citep{marijan2022industry,garousi2019characterizing}.

At the same time, \textbf{RQ4} highlights an important measurement issue.
Commits, contributors, stars, forks, and watchers provide observable
repository-level indicators of activity and community engagement, but they do
not directly establish CE maturity, production use, or resilience
improvement. Future empirical work should therefore examine whether and under
what conditions these repository indicators correspond to stronger evidence
of CE adoption and operational practice.

\begin{tcolorbox}[colback=blue!3, colframe=black!50,
boxrule=0.8pt, arc=1pt, left=4pt, right=4pt, top=2pt, bottom=2pt]

\faFlask\ \textbf{Implication for Researchers \#2.}

Researchers should investigate the factors behind the comparatively limited
representation of academic-owned CE repositories and validate the extent to
which repository activity and engagement indicators correspond to CE
maturity, operational use, or resilience outcomes.

\end{tcolorbox}

Finally, \textbf{RQ5} identifies a substantial imbalance in the fault types
represented in the studied repositories. The dominance of network and
instance-termination faults raises questions about why application-level
failures are comparatively uncommon and whether predominantly
infrastructure-oriented experiments provide sufficiently broad evidence about
system resilience. Future research should investigate the technical and
organizational factors behind this imbalance and evaluate whether broader
fault portfolios expose weaknesses that infrastructure-focused experiments
may not reveal.

\begin{tcolorbox}[colback=blue!3, colframe=black!50,
boxrule=0.8pt, arc=1pt, left=4pt, right=4pt, top=2pt, bottom=2pt]

\faFlask\ \textbf{Implication for Researchers \#3.}

Researchers should investigate why application-level faults are
underrepresented in repository-visible CE practice and evaluate whether
broader fault portfolios provide different or more comprehensive evidence of
system resilience.

\end{tcolorbox}

\textcolor{blue}{
\paragraph{Emerging research direction.}
Beyond the implications derived directly from RQ1-RQ5, AI-assisted software engineering represents an emerging direction for CE. As AI models increasingly support fault and vulnerability detection, they may also assist CE activities such as hypothesis formulation, failure-scenario generation, and experiment design. Early prototypes such as ChaosEater\footnote{\url{https://github.com/ntt-dkiku/chaos-eater.}} illustrate this direction. Future studies should therefore evaluate their effectiveness, safety, and applicability in real-world resilience engineering.}

\section{Threats to Validity}
\label{sec:threats}
This section examines potential threats to the validity of our research and its findings.

\paragraph{Internal Validity}
This concerns the soundness of the study procedures and control of
potential bias~\cite{wohlin2012experimentation}. Manual repository and purpose
classification may introduce researcher subjectivity. We mitigated this through
predefined criteria, initial classification by the first author, review by the
second author, and consensus resolution of ambiguous cases. False positives from
automated searches were removed through manual inspection of repository
documentation, metadata, configuration files, and source code. For RQ5,
undocumented or implicit CE usage may still lead to under- or overestimation;
we reduced this risk by examining configuration files, scripts, and relevant
code paths. GitHub API limitations and incomplete or non-English documentation
may nevertheless have excluded or obscured some repository evidence.

\paragraph{Construct Validity}

This concerns whether the operational measures adequately
represent the intended concepts~\citep{wohlin2012experimentation}. First,
the presence of a CE tool in a repository does not necessarily indicate active
use. We mitigated this by manually validating candidate repositories against
configuration files, documentation, and source code, retaining only verified
CE-adopting repositories. Second, the GitHub metrics used in RQ4 (e.g.,
commits, contributors, releases, stars, forks, and watchers) capture observable
project activity and prominence rather than software quality or industrial
impact. Finally, RQ5 captures only explicitly observable CE usage, so
undocumented or implicit capabilities may remain undetected.

\paragraph{External Validity}

This concerns the extent to which findings generalize beyond the
study context~\cite{wohlin2012experimentation}. First, our analysis is limited
to public GitHub repositories and therefore excludes other platforms, such as
GitLab and Bitbucket, as well as private and proprietary repositories
~\citep{kalliamvakou2016depth,tutko2022software}. We mitigated this threat by
using GitHub, a widely adopted source in repository-mining research, but the
findings should still be interpreted as characterizing the studied public
repositories rather than CE adoption across the software industry. Second, focusing on the ten most prominent CE tools may favor active and
well-maintained projects while underrepresenting niche or inactive tools. We
mitigated this by deriving the tool set from both prior literature and GitHub
activity and cross-checking it with previous studies
~\citep{sanchez2022mutation,ustunboyacioglu2024data}. Tools such as Chaos
Engine, Namazu, Simian Army, and ChaoSlingr were excluded because of archival
status or limited development
~\citep{ThalesGroup_ChaosEngine,OSRG_Namazu,Netflix_SimianArmy,
Optum_ChaoSlingr}. Nevertheless, the results may not generalize to excluded
tools or platforms; replication in broader settings would strengthen external
validity.

\paragraph{Conclusion Validity}
This concerns the degree to which the study’s conclusions are reasonably based on the data analyzed~\cite{wohlin2012experimentation}.
Some of our findings, including RQ1 adoption trends, are based on counts of CE tool usage over time. Without normalization against overall GitHub growth, increases may reflect platform expansion rather than greater CE adoption~\citep{cosentino2016findings,borges2016understanding}. Future studies should normalize these trends using total repository counts to assess proportional growth in CE usage~\citep{kalliamvakou2014promises}.

\section{Conclusion}
\label{sec:conclusion}

This paper examines the adoption of chaos engineering tools in open-source GitHub projects. We identified the ten most used tools, manually removed false positives, and analyzed 1,275 repositories to examine adoption, activity, and usage. Adoption rose sharply after 2018. Most repositories support development, with fewer used for learning, teaching, or research. Industry represents the largest ownership category, closely followed by personal repositories, while public and academic ownership is less common. Repository activity varies widely, although many repositories show limited commit and contributor activity. Tools support different fault types: Chaos Mesh spans instance, network, resource, and application faults; Toxiproxy targets network faults; and Chaoskube and Chaos Monkey focus on instance termination. Across repositories, network faults are the most common scenario, followed by instance termination, resource stress, and application faults. In future work, we plan to investigate the use of CE principles and practices to evaluate and enhance the robustness of multi-agent systems powered by large language models (LLM-MAS). We aim to develop a fault taxonomy for LLM-MAS through a literature review, GitHub mining, and interviews with practitioners. We will also study the recovery strategies for each fault by extending existing CE tools to inject and observe faults. Furthermore, future work could investigate integrating CE tools with AI-driven monitoring systems to automatically identify and respond to faults in real time, thereby improving the resilience and self-healing capabilities of software systems.

\section*{CRediT authorship contribution statement}

\textbf{Joshua Owotogbe}: Data collection, Formal analysis, Visualization, Writing – original draft.  
\textbf{Indika Kumara}: Validation, Writing – review \& editing.  
\textbf{Dario Di Nucci}: Writing – review \& editing.  
\textbf{Damian Andrew Tamburri}: Writing – review \& editing.  
\textbf{Willem-Jan van den Heuvel}: Supervision, Writing – review \& editing.

\section*{Declaration of competing interest}
The authors declare no competing financial or personal interests.

\section*{Funding}
This research received no external funding.

\section*{Data availability}
The online appendix\footnote{\url{https://zenodo.org/records/22340305}} provides the scraping and mining code, repository details, and the dataset used to answer the research questions.

\section*{Ethical approval}

Not applicable.

\section*{Informed consent}

Not applicable.

\section*{Clinical trial number}

Not applicable.

\bibliographystyle{cas-model2-names}


\bibliography{references}

\end{document}